\documentclass[fleqn,usenatbib]{mnras}

\usepackage{newtxtext,newtxmath}
\usepackage[T1]{fontenc}
\usepackage{ae,aecompl}

\usepackage{graphicx}	% Including figure files
\usepackage{amsmath}	% Advanced maths commands
\usepackage{amssymb}	% Extra maths symbols
\usepackage{gensymb}
\usepackage{threeparttable}
\usepackage{tablefootnote}

\title[]{Rapid Black Hole Growth at the Dawn of the Universe: A Super-Eddington Quasar at z=6.6}

\author[J.-J. Tang et al.]{
Ji-Jia Tang,$^{1,2,3}$\thanks{E-mail: jjtang@asiaa.sinica.edu.tw, ji-jia.tang@anu.edu.au}
Tomotsugu Goto,$^{4}$
Youichi Ohyama,$^{1}$
Chichuan Jin,$^{5}$
Chris Done,$^{6}$
\newauthor
Ting-Yi Lu,$^{4}$
Tetsuya Hashimoto,$^{4}$
Ece Kilerci Eser,$^{4,7}$
Chia-Ying Chiang,$^{4}$
\newauthor
and Seong Jin Kim$^{4}$
\\
$^{1}$Institute of Astronomy and Astrophysics, Academia Sinica, No.1, Sec. 4, Roosevelt Rd., Taipei 10617, Taiwan\\
$^{2}$Graduate Institute of Astrophysics, National Taiwan University, No.1 Sec.4 Roosevelt Rd., Taipei 10617, Taiwan\\
$^{3}$Research School of Astronomy and Astrophysics, Australian National University, Cotter Road, Weston Creek, ACT 2611, Australia\\
$^{4}$Institute of Astronomy, National Tsing Hua University, 101, Sec. 2, Kuang Fu Rd., Hsinchu 30013, Taiwan\\
$^{5}$National Astronomical Observatories, Chinese Academy of Sciences, A20 Datun Road, Beijing 100101, China\\
$^{6}$Centre for Extragalactic Astronomy, Department of Physics, University of Durham, South Road, Durham DH1 3LE, UK\\
$^{7}$ Istanbul University, Science Faculty, Department of Astronomy and Space Sciences, Beyaz{\i}t, 34119, Istanbul, Turkey 
}

\date{Accepted XXX. Received YYY; in original form ZZZ}

\pubyear{2019}

\begin{document}
\label{firstpage}
\pagerange{\pageref{firstpage}--\pageref{lastpage}}
\maketitle

\begin{abstract}
We present the analysis of a new near-infrared (NIR) spectrum of a recently discovered $z=6.621$ quasar PSO J006+39 in an attempt to explore the early growth of supermassive black holes (SMBHs). 
This NIR (rest-frame ultraviolet, UV) spectrum shows blue continuum slope and rich metal emission lines in addition to Ly$\alpha$ line.
We utilize the $\ion{Mg}{II}$ line width and the rest frame luminosity $L_\text{3000\AA}$ to find the mass of SMBH ($M_\text{BH}$) to be $\sim 10^8 M_\odot$,
making this one of the lowest mass quasars at high redshift.
The power-law slope index ($\alpha_\lambda$) of the continuum emission is $-2.94\pm0.03$,
significantly bluer than the slope of $\alpha_\lambda=-7/3$ predicted from standard thin disc models.
We fit the spectral energy distribution (SED) using a model which can fit local SMBHs, 
which includes warm and hot Comptonisation powered by the accretion flow as well as an outer standard disc. 
The result shows that the very blue slope is probably produced by a small radial ($\sim230$ gravitational radius, $R_\text{g}$) extent of the standard accretion disc. 
All plausible SED models require that the source is super-Eddington ($L_\text{bol}/L_\text{Edd} \gtrsim 9$), 
so the apparently small disc may simply be the inner funnel of a puffed up flow,
and clearly the SMBH in this quasar is in a rapid growth phase. 
We also utilize the rest-frame UV emission lines to probe the chemical abundance in the broad line region (BLR) of this quasar.
We find that this quasar has super solar metallicity through photoionization model calculations.
\end{abstract}


%%%%%%%%%%%%%%%%%%%%%%%%%%%%%%%%%%%%%%%%%%%%%%%%%%

%%%%%%%%%%%%%%%%% BODY OF PAPER %%%%%%%%%%%%%%%%%%

\defcitealias{2017ApJ...849...91M}{M17}
\defcitealias{2006A&A...447..157N}{N06}
\defcitealias{1973A&A....24..337S}{SS73}

\section{Introduction}
\label{sec:intro}

Quasars are powered by the accretion disc that surrounds the supermassive black hole (SMBH) at the center of the host galaxies.
The thermal emission from accretion disc dominates in the rest-frame ultraviolet (UV) to optical.
The standard thin disc of \citet{1973A&A....24..337S} (hereafter \citetalias{1973A&A....24..337S}) predicts that the continuum emission has a power-law shape by assuming a multi-temperature blackbody model.
The power-law shape where $f_\nu \propto \nu^{\alpha_\nu}$ ($f_\lambda \propto \lambda^{\alpha_\lambda}$) has a slope of $\alpha_\nu=+1/3$ ($\alpha_\lambda=-7/3$).
However, the observed continuum slopes are usually much redder.
\citet{2001AJ....122..549V} created a composite spectrum from $z<5$ quasars in Sloan Digital Sky Survey (SDSS) and found the continuum slope is $\alpha_\lambda=-1.5$.
Another composite quasar spectrum covering the rest frame UV to near-infrared (NIR) showed that the continuum slope is $\alpha_\lambda=-1.7$ \citep{2016A&A...585A..87S}.
\citet{2016ApJ...824...38X} showed the average continuum slope of quasars from SDSS DR7 are $\alpha_{\lambda}=-1.64$ and $\alpha_{\lambda}=-1.49$ for near UV and far UV with similar dispersions $\sim 0.5$,
but a small portion of quasars in their samples have a bluer slope of $\alpha_{\lambda}>-7/3$.
\citet{2017ApJ...849...91M} (hereafter \citetalias{2017ApJ...849...91M}) analysed the continuum slope of 15 quasars at $z \gtrsim 6.5$, including our target J006.1240+39.2219 (hereafter PSO J006+39) in this study.
Unfortunately, they did not cover wide enough NIR spectrum to obtain precise continuum slopes for PSO J006+39 and other two quasars.
The average slope they found for 12 quasars at $z \gtrsim 6.5$ with NIR spectrum is $\alpha_\lambda=-1.2\pm0.4$.
Based on our current knowledge,
only a few high-redshift quasars have continuum slopes bluer than \citetalias{1973A&A....24..337S} prediction of $\alpha_\lambda=-7/3$.
\citet{2010A&A...523A..85G} showed that 7 out of 33 quasars in the range of $3.9 \leq z \leq 6.4$ in their samples could have very blue continuum slope ($\alpha_\lambda \leq -7/3$) after extinction correction.
\citet{2007AJ....134.1150J} reported that two $z\sim6$ quasars, SDSS J1306+0356 and SDSS J1030+0524, have $\alpha_\lambda \sim -2.5$.
However, \citet{2011ApJ...739...56D} showed that SDSS J1030+0524 is not such blue after they reanalysed the spectrum with different fitting wavelength range.
In the end, only one quasar is confirmed to be bluer than $-7/3$ out of 21 $z > 6$ quasars.

Several physical parameters of quasars affect the continuum slope.
Dust reddening is a conventional explanation to the overall redder slope \citep[e.g.][]{2007ApJ...668..682D,2016ApJ...824...38X,2016ApJ...818L...1S}.
The observational evidence suggests that black hole mass ($M_\text{BH}$) does not affect the slope \citep[e.g.][]{2007ApJ...668..682D}.
In theory, higher $M_\text{BH}$ will shift the peak of disc emission to lower frequency \citep[e.g.][]{2018A&A...612A..59C}.
Besides, the lower inclination angle of the disc and the higher black hole spin will both make the slope bluer \citep[e.g.][]{2016ApJ...818L...1S},
but they are observationally hard to confirm.
Figure 5 of \citet{2018A&A...612A..59C} showed that increasing black hole spin and accretion rate ($\dot{M}$) shift (in different directions) the disc emission peak to brighter luminosities and higher frequencies.
These parameters will affect the continuum slope in a complicated way.
Besides, the corona of the disc can create a soft X-ray excess \citep[e.g.][]{2012MNRAS.420.1848D, 2017MNRAS.468.1442H}, 
which may also extend to the UV regime and make the slope bluer.

The existence of quasars at early universe indicates that the central SMBHs can grow in a short timescale.
There are more than hundreds of quasars known at $z \gtrsim 6$ universe \citep[e.g.][]{2016ApJS..227...11B}.
Most of them are discovered in this decade, 
including the highest redshift quasar at $z=7.54$ known so far \citep{2018Natur.553..473B}.
\citetalias{2017ApJ...849...91M} estimated $M_\text{BH}$ of 11 $z \gtrsim 6.5$ quasars and found $M_\text{BH} \gtrsim 10^8 M_\odot$ for all of them.
\citet{2015Natur.518..512W} even found a SMBH with $M_\text{BH} \sim 10^{10} M_\odot$ at $z=6.30$.
These SMBHs must have been growing at the most efficient rate to become such massive objects even at the very early universe.
Several models aim at solving the issue of how SMBHs can grow so fast (see references in \citetalias{2017ApJ...849...91M}).
Providing more $M_\text{BH}$ measurements for $z > 6.5$ quasars is useful to constrain these models.
The estimation of $M_\text{BH}$ for high redshift quasars is usually done by adopting the empirical relations \cite[e.g.][]{2011ApJ...739...56D,2014ApJ...790..145D,2017ApJ...849...91M}.
These empirical relations are derived from the broad emission line-width and the continuum luminosity \citep[e.g.][]{2009ApJ...699..800V} by assuming that the strong correlation between continuum luminosity and the size of the broad-line region (BLR) in the low redshift active galactic nuclei (AGNs) \citep[e.g.][]{2013ApJ...767..149B} is independent of redshift.
The $M_\text{BH}$ of the quasar can be further used to obtain Eddington luminosity ($L_\text{Edd}$).
Those SMBHs in super-Eddington ($L_\text{bol}/L_\text{Edd} > 1$) high redshift quasars found by \citetalias{2017ApJ...849...91M} are most likely to grow rapidly.

It is expected that the chemical abundances are lower for higher redshifted quasars.
One typical method to determine the chemical abundance for high redshift quasars is using Fe/Mg abundance ratio.
This ratio is a good tracer to the star formation history across cosmic time.
The $\alpha$-elements like Mg is primarily formed from Type II supernovae (SNe II).
Fe is most likely produced by Type Ia supernovae (SNe Ia), 
which take $\sim$ 1Gy longer time to form comparing to SNe II \cite[e.g.][]{1986A&A...154..279M}.
This time scale is approximated to the age of the universe at $z\sim 6$;
therefore, the Fe/Mg abundance ratio is expected to be lower at $z\gtrsim6$.
However,  
there is no concrete observational evidence supporting this expectation so far \citep[e.g.][]{2002ApJ...565...63I,2003ApJ...594L..95B,2003ApJ...596..817D,2003ApJ...587L..67F,2003ApJ...596L.155M,2004ApJ...614...69I,2007AJ....134.1150J,2011ApJ...739...56D}.
Although \citetalias{2017ApJ...849...91M} suggested that \ion{Fe}{II}/\ion{Mg}{II} flux ratio, which can be regarded as Fe/Mg abundance ratio, of $z\gtrsim6$ quasars show systematically lower value compared to low redshift quasars,
they cannot confirm this due to large uncertainties in their results.
Another method to determine the chemical abundance is estimating metallicity by utilizing several line ratios such as \ion{N}{V}$\lambda1240$/\ion{He}{II}$\lambda1640$, \ion{N}{V}$\lambda1240$/\ion{C}{IV}$\lambda1549$, and  (\ion{Si}{IV}$\lambda1398$+\ion{O}{IV]}$\lambda1402$)/\ion{C}{IV}$\lambda1549$ \citep[e.g.][]{2006A&A...447..157N, 2007AJ....134.1150J, 2011A&A...527A.100M, 2018MNRAS.480..345X}.
This method requires a good signal-to-noise ratio (SNR) spectrum to detect those faint lines.
\citet{2009A&A...494L..25J} reported the metallicities of quasars up to $z \sim 6.4$ by applying this method,
and found no redshift evolution.
Note that \ion{Fe}{II}/\ion{Mg}{II} ratio is not likely to be a good indicator of metallicity at $z>6$,
because Fe/Mg abundance ratio is expected to be constant before beginning of chemical enrichment by SNe Ia \citep[e.g.][]{1997ARA&A..35..503M}. 

In this paper, we examine a new NIR spectrum of the quasar PSO J006+39 to obtain the continuum slope, $M_\text{BH}$, and metallicity properties in detail. 
It was discovered from Panoramic Survey Telescope \& Rapid Response System 1 \citep[Pan-STARRS1 or PS1;][]{2002SPIE.4836..154K,2010SPIE.7733E..0EK} and confirmed by optical spectrum from Subaru Telescope \citep{2017MNRAS.466.4568T}. 
This quasar is bright ($m_{y_\text{PS1}}=20.06\pm0.07$) and has strong emission lines.
It is an ideal target to investigate the growth rate of SMBH and the metallicity of the quasar.
We adopt the redshift measurement from \citetalias{2017ApJ...849...91M} for this quasar.
They reported the systemic redshift of this quasar from [\ion{C}{II}] 158 $\mu$m line observation and found $z_\text{[\ion{C}{II}]}=6.621 \pm 0.002$. 
This is currently the most precise redshift measurement of PSO J006+39.

The structure of this paper is organised as follows:
In Sec.~\ref{sec:obdr} we describe the Gemini observation and the data reduction for the NIR spectrum.
In Sec.~\ref{sec:fit} we describe the spectral fitting for the spectrum.
In Sec.~\ref{sec:dar} we show the results on the continuum slope and the absolute magnitude (Sec.~\ref{sec:slope}),
the SMBH mass and the Eddington ratio (Sec.~\ref{sec:mbh}),
the SED modeling (Sec.~\ref{sec:sed}),
and the chemical abundance in the BLR (Sec.~\ref{sec:metal}).
In Sec.~\ref{sec:dis} we discuss implication of the SED modeling (Sec.~\ref{sec:impslope}) and the metallicity comparison with other quasars (Sec.~\ref{sec:commet}).
In Sec.~\ref{sec:con} we summarise our findings.
We use a $\Lambda$CDM cosmology with $H_0=70$ km s$^{-1}$ Mpc$^{-1}$, $\Omega_m=0.3$, and $\Omega_\Lambda=0.7$.

\section{Observations and Data Reduction}
\label{sec:obdr}

\begin{figure*}
	\includegraphics[width=\textwidth]{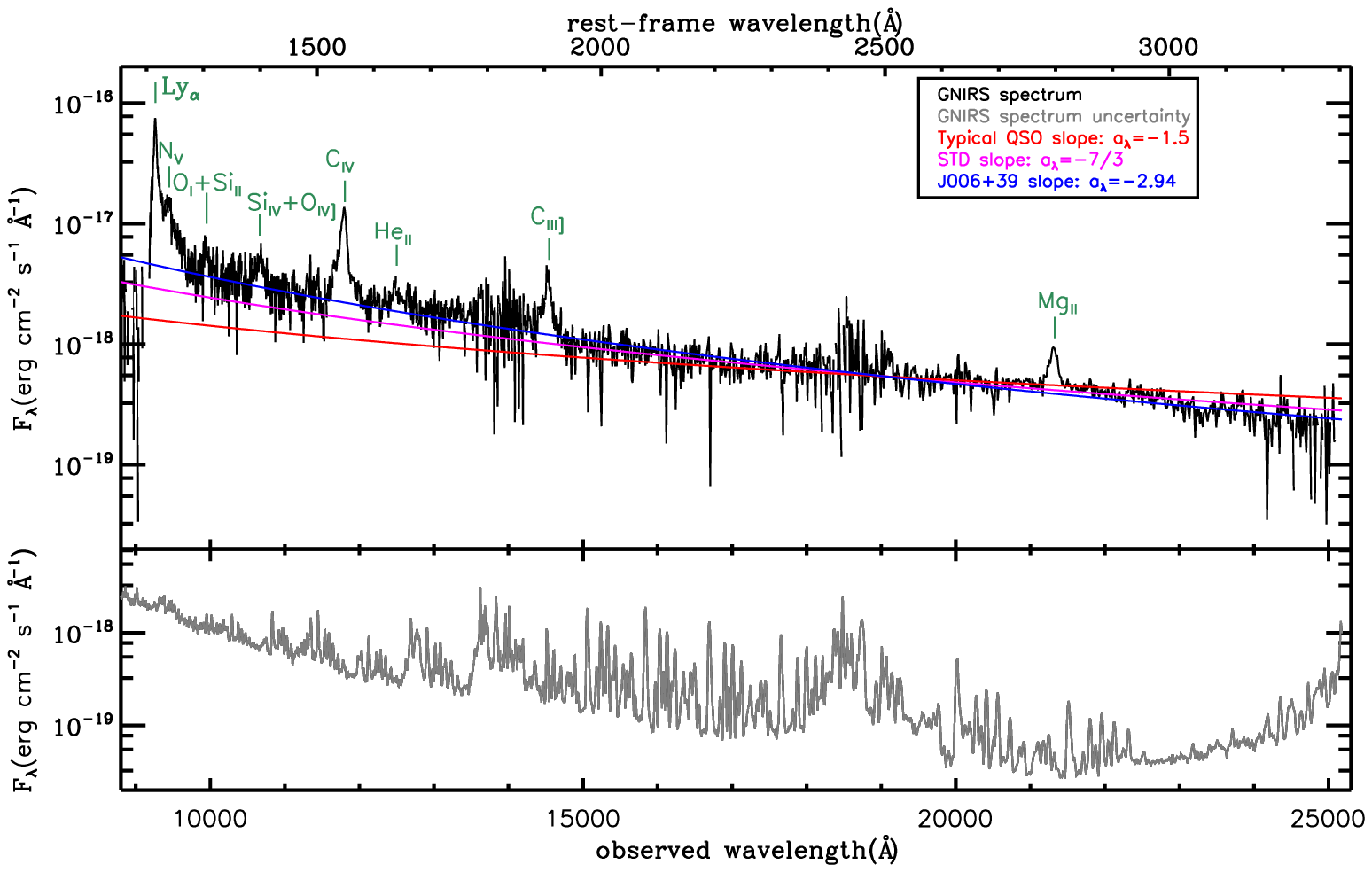}
	\caption{GNIRS spectrum of PSO J006+39.
	The black and gray lines show the spectrum and its error, respectively.
	Blue, magenta and red slopes are the continuum slopes of PSO J006+39, standard thin disc model, and typical quasars \citep{2001AJ....122..549V}, respectively.
	Detected emission lines are labeled in green lines.
	The spectrum and its error are shown in 2 binning for demonstration purpose.\protect\\
	(A colour version of this figure is available in the online journal.)}
	\label{fig:QSOspec}
\end{figure*}

The NIR spectrum of PSO J006+39 was taken with Gemini Near Infra-Red Spectrograph (GNIRS) with cross-dispersed (XD) mode on 2016, Aug. 3rd, 5th, 8th, and 9th (PI: Tomotsugu Goto). 
The slit-width was $1''$ with a 32/mm grating. 
The seeing during the observation ranged from $0''.5$ to $0''.75$. 
Each frame was exposed for 300 sec and the total exposure time was 4 hrs. 
We applied the ABBA pattern nodding strategy among the frames.
We used the Gemini {\sc iraf} package to reduce the spectra. 
We followed the standard procedure with some corrections to optimize the result.
The standard procedure includes the method of concatenated different echelle orders in the XD spectrum.
We concatenated the standard star spectrum by assuming it is in general a black body spectrum,
then scaled each order of the quasar XD spectrum with respect to corresponding order of the standard star spectrum.
This method ensures that the shape of the quasar XD spectrum is reliable. 
We hope that the future near-IR photometric observations in J, H, K bands can confirm it
The corrections include images shifting correction due to instrumental flexure and strong hydrogen absorption lines correction in the spectrum of the telluric standard star.
At last, we scaled the quasar NIR spectrum according to the PS1 $y$-band magnitude $m_{y_\text{PS1}} = 20.06 \pm 0.07$. 
The calibrated spectrum is shown in Fig.~\ref{fig:QSOspec}.

\section{Spectral Analysis}
\label{sec:fit}

\begin{figure*}
	\includegraphics[width=\textwidth]{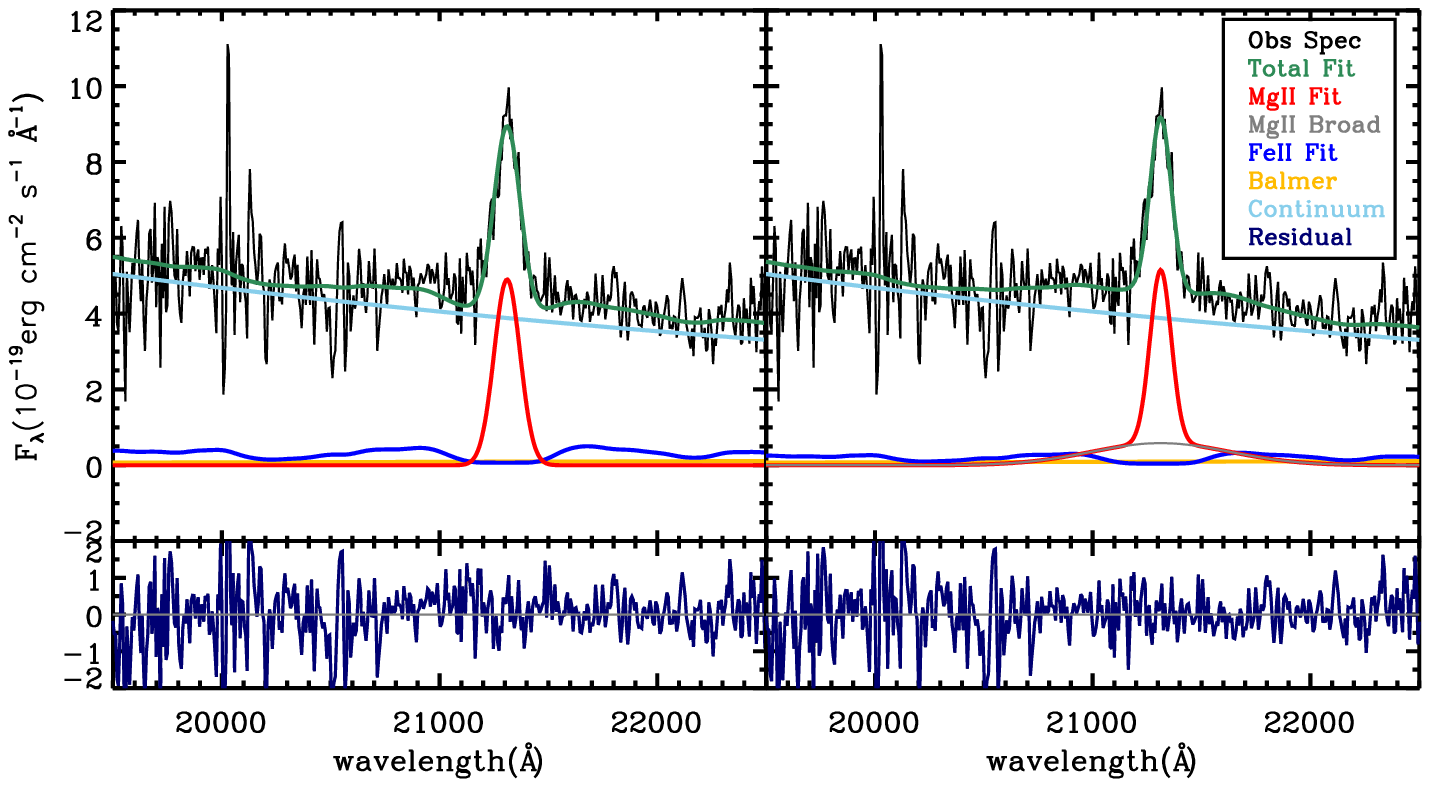}
	\caption{Spectrum around \ion{Mg}{II} and out best fit results.
	The left panel shows the result of the single Gaussian function fit while the right panel shows the double Gaussian function fit for \ion{Mg}{II} line.
	The black line is the observed spectrum (Fig.~\ref{fig:QSOspec}).
	The rest of the fitting components are \ion{Mg}{II} line (red), \ion{Fe}{II} pseudo-continuum (blue), Balmer pseudo-continuum (gold), power-law continuum (sky blue), total (green), and residual (navy).\protect\\
	(A colour version of this figure is available in the online journal.)}
	\label{fig:mg2fe2fit}
\end{figure*}

To obtain the continuum slope, $M_\text{BH}$, and the \ion{Fe}{II}/\ion{Mg}{II} ratio, we fit the NIR spectrum of this quasar.
We follow the same method used by \citetalias{2017ApJ...849...91M} and apply $\chi^2$-minimization technique in the following way.
We first fit a single power-law continuum in the following rest-frame wavelength windows to avoid emission lines.
[1285--1295; 1315--1325; 1340--1375; 1425--1470; 1680--1710; 1975--2050; 2150--2250; and 2950--2990] \AA\AA~\citep{2009ApJ...699..782D}.  
The power-law equation is: 
\begin{equation}
	F_\lambda=F_0 (\frac{\lambda}{2500\text{\AA}})^{\alpha_\lambda}
	\label{eq:powerlaw}
\end{equation}
Then we consider the Balmer pseudo-continuum from Equation (7) in \citet{1982ApJ...255...25G} and assume that it is 10\% of the global power-law continuum at $\lambda_\text{rest}=3675 \text{\AA}$.
After determining these two components, 
we fit the pseudo-continuum \ion{Fe}{II} emission and the \ion{Mg}{II} emission line within the rest-frame wavelength composite window 2100--3200 \AA\AA.
We adopt the empirical pseudo-continuum \ion{Fe}{II} emission template from \citet{2001ApJS..134....1V} and apply the FWHM $=15~$\AA~for the \ion{Fe}{II} as suggested by \citet{2011ApJ...739...56D}.
Finally, we fit the \ion{Mg}{II} emission line with a single Gaussian function.
The result of this method is shown in the left panel of Fig.~\ref{fig:mg2fe2fit} and Table~\ref{tab:lines}.

\begin{figure}
	\includegraphics[width=\columnwidth]{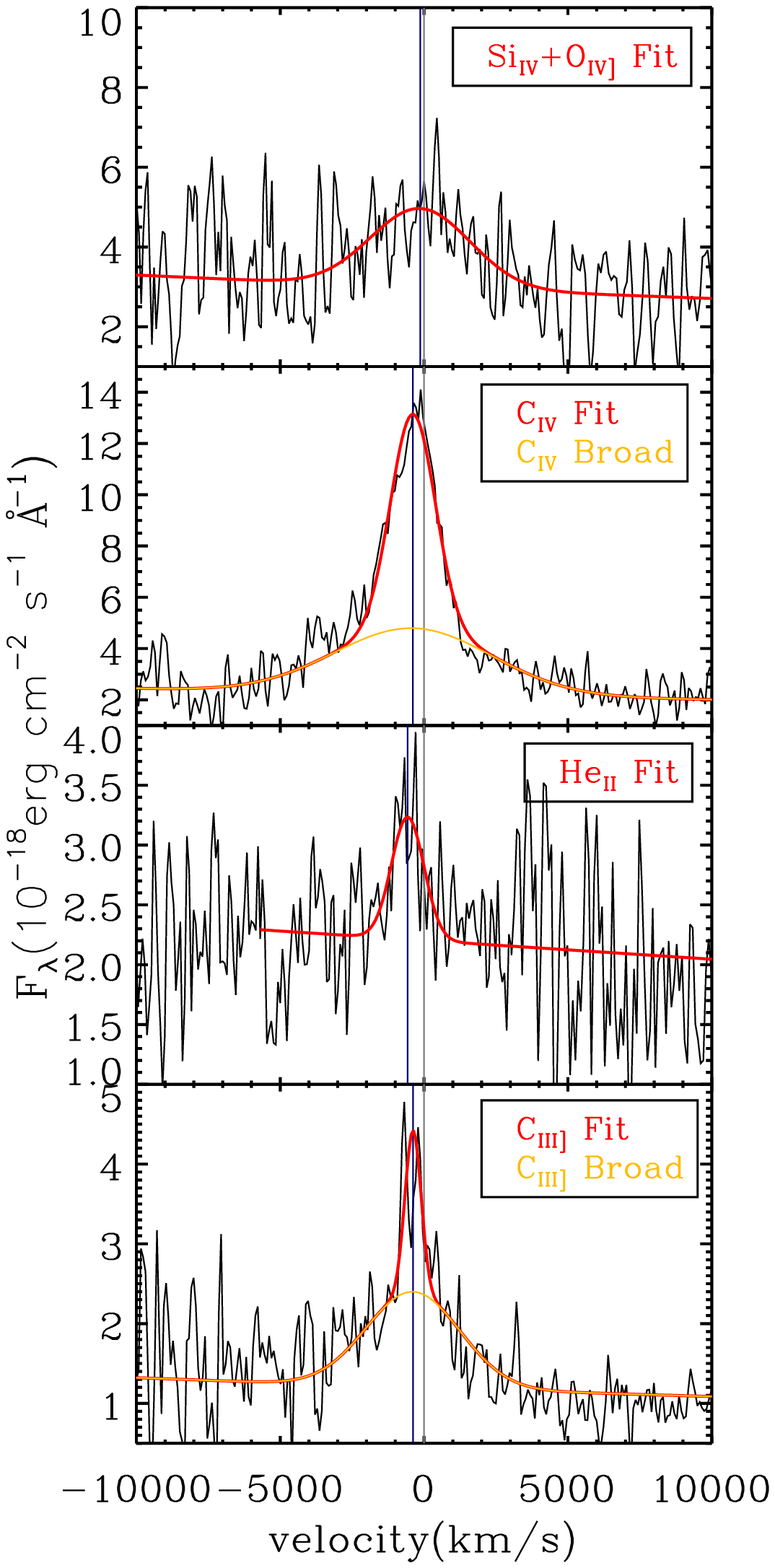}
	\caption{Fitting results of emission lines.
	From the top panel to the bottom panel, these panels show \ion{Si}{IV}+\ion{O}{IV]}, \ion{C}{IV}, \ion{He}{II}, and \ion{C}{III]} lines (black) and their fitting results (red). 
	The fitting for \ion{Si}{IV}+\ion{O}{IV]} and  \ion{He}{II} lines used a single Gaussian function and local continuum fit while the \ion{C}{IV} and \ion{C}{III]} used a double Gaussian function with the continuum slope in Fig.~\ref{fig:QSOspec}.
	The gray vertical lines indicate the systemic redshift from [\ion{C}{II}] line while the navy vertical lines indicate the fitted center of each line.\protect\\
	(A colour version of this figure is available in the online journal.)}
	\label{fig:linesfit}
\end{figure}

Since there are some residuals around the \ion{Mg}{II} emission, 
we also try a double Gaussian function with a fixed center for both Gaussian functions for \ion{Mg}{II} line (Fig.~\ref{fig:mg2fe2fit} right).
The double Gaussian function fit results in smaller residuals compared to the single Gaussian function fit. 
In fact, all other strong emission lines (Ly$\alpha$, \ion{C}{IV}, and \ion{C}{III]}) require double Gaussian functions to produce a better fit (see \citealp{2017MNRAS.466.4568T} for Ly$\alpha$, and Fig.~\ref{fig:linesfit} for \ion{C}{IV} and \ion{C}{III]} lines).

We also identify faint emission lines (\ion{Si}{IV}+\ion{O}{IV]} and \ion{He}{II}) and fit them with single Gaussian functions (Fig.~\ref{fig:linesfit}).
As for \ion{N}{V} and \ion{O}{I}+\ion{Si}{II} lines,
they can not be well fitted with Gaussian profiles in this NIR spectrum.
Therefore, we utilize the optical spectrum of PSO J006+39 (Lu, T.-Y. et al. in preparation) to obtain the line luminosity.
We present the line luminosity, velocity shift, FWHM, and EW of all detected lines in Table~\ref{tab:lines}.
The uncertainties shown in Table~\ref{tab:lines} are statistical errors.
The result based on the double Gaussian fitting method for \ion{Mg}{II} is utilized in Sec.~\ref{sec:slope}, Sec.~\ref{sec:mbh}, and Sec.~\ref{sec:metal}.
It is found that in general,
the \ion{C}{IV} lines of high-redshifted quasars tend to show large blueshift ($\sim$ 1000 km s$^{-1}$) compared to their host galaxies \cite[e.g.][]{2016ApJ...816...37V,2017ApJ...849...91M}.
This \ion{C}{IV} blueshift can be explained by the existence of outflow in BLR \citep[e.g.][]{2004ApJ...611..125L}.
For PSO J006+39, it only shows a relatively small amount of \ion{C}{IV} blueshift ($-387.8 \pm 0.1$ km s$^{-1}$).
It is also shown that the \ion{C}{IV} blueshift is correlated with the equivalent-width (EW) of \ion{C}{IV} \citep{2011AJ....141..167R}.
Note that the \ion{C}{IV} EW in Table~\ref{tab:lines} is twice as large as other $z>6.5$ quasars in \citetalias{2017ApJ...849...91M}.

\begin{table*}
	\centering
	\caption{The emission line properties.}
	\label{tab:lines}
	\begin{threeparttable}
	\begin{tabular}{lccccc} 
		\hline
		Lines & $\lambda_\text{rest}$ (\AA) & Luminosity (10$^{44}$ erg s$^{-1}$) & Velocity shift (km s$^{-1}$) & Width (km s$^{-1}$) & EW (\AA) \\
		\hline
		\ion{N}{V} & 1240.14 & $1.79 \pm 0.06$ \tnote{a} &  &  & \\
		\ion{O}{I}+\ion{Si}{II} & 1305.59 & $1.54 \pm 0.16$ \tnote{a} & & & \\
		\ion{Si}{IV}+\ion{O}{IV]} & 1399.41 & $1.75 \pm 0.16$ & $-132 \pm 164$ & $4072 \pm 330$ & $15.1\pm1.7$ \\
		\ion{C}{IV} & 1549.06 & $7.15 \pm 0.45$ & $-388 \pm 80$ & $1820 \pm 49$ & $84.4 \pm 2.2$ \\
		$\ion{C}{IV}_\text{broad}$ & 1549.06 &  &  & $6456 \pm 7$ &  \\
		\ion{He}{II} & 1640.42 & $0.67 \pm 0.05$ & $-569\pm 104$ & $1333 \pm169$ & $9.3 \pm 7.2 $ \\
		\ion{C}{III]} & 1908.73 & $2.07 \pm 0.36$ & $-381 \pm 85$ & $662 \pm 79$ & $43.5 \pm 3.8$ \\
		$\ion{C}{III]}_\text{broad}$ & 1908.73 &  &  & $3847 \pm 50$ &  \\
		$\ion{Mg}{II}_\text{double}$ & 2798.75 & $0.51 \pm 0.07$ & $-237 \pm 83$ & $1620 \pm 73$ & $34.6 \pm 7.8$ \\
		$\ion{Mg}{II}_\text{double,broad}$ & 2798.75 &  &  & $10550 \pm 83$ &  \\
		$\ion{Mg}{II}_\text{single}$ & 2798.75 & $0.37 \pm 0.01$ & $-265 \pm 83$ & $1972 \pm 55$ & $24.7 \pm 7.6$ \\
		\hline
	\end{tabular}
	\begin{tablenotes}
		\item[a] Values are taken from Lu, T.-Y. et al. in preparation.
	 \end{tablenotes}	
	 \end{threeparttable}
\end{table*}

\section{Results}
\label{sec:dar}

\subsection{Continuum and absolute magnitude, \texorpdfstring{$M_{1450}$}{TEXT}}
\label{sec:slope}

\begin{figure}
	\includegraphics[width=\columnwidth]{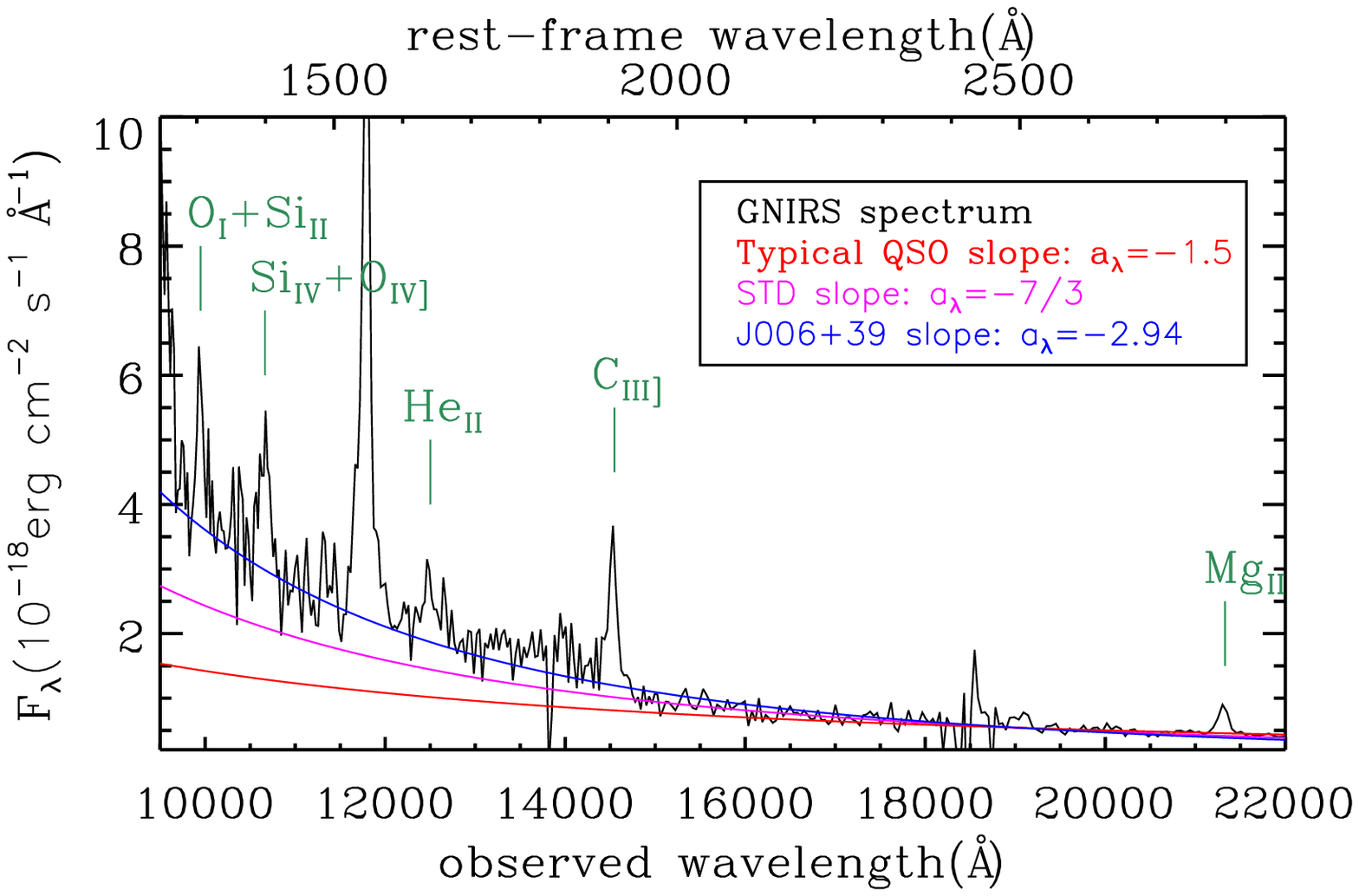}
	\caption{The zoom in spectrum of Fig.~\ref{fig:QSOspec}.
	The lines and labels are the same as in Fig.~\ref{fig:QSOspec}.
	The spectrum is shown in 8 binning for demonstration purpose.\protect\\
	(A colour version of this figure is available in the online journal.)}
	\label{fig:binQSOspec}
\end{figure}

We show the power-law continuum slope and absolute magnitude here.
From the fitting in Sec.~\ref{sec:fit}, 
the power-law parameters in Eq.~\ref{eq:powerlaw} are $\alpha_\lambda=-2.94 \pm 0.03$ ($\alpha_\nu=0.94 \pm 0.03$) and $F_0=5.40 \pm 0.05 \times 10^{-19}$ erg s$^{-1}$ cm$^{-2} \text{\AA}^{-1}$.
This slope is much bluer compared to any other known quasars at $z \gtrsim 6$ ($\alpha_\lambda \sim -1.5$).
We further verify this slope by 4, 8, and 16 binnings of the spectrum before continuum slope fitting.
All fitting results show consistent slopes.
We show the case of 8 binning in Fig.~\ref{fig:binQSOspec} as an example.
We calculate the apparent magnitude at rest-frame 1450\AA~as $m_\text{1450}=21.30 \pm 0.07$ by extrapolation based on the fit.
This corresponds to absolute magnitude $M_\text{1450}=-25.55 \pm 0.07$, 
which is slightly fainter than the value reported in \citet{2017MNRAS.466.4568T} ($-25.94$).

\subsection{SMBH mass and Eddington ratio}
\label{sec:mbh}

To determine $M_\text{BH}$, we adopt the empirical relation provided by \citet{2009ApJ...699..800V}:
\begin{equation}
	\text{log}(\frac{M_\text{BH}}{M_\odot})=6.86+2\times \text{log}(\frac{\text{FWHM}}{10^3\text{km s}^{-1}})+0.5\times \text{log}(\frac{\lambda L_{\lambda,3000}}{10^{44}\text{erg s}^{-1}}).
	\label{eq:mbheq}
\end{equation}
This relation is only applicable to the \ion{Mg}{II} emission line.
The method of obtaining the FWHM of the \ion{Mg}{II} is described in Sec.~\ref{sec:fit}.
The 1$\sigma$ scatter of the zero-point of this relation is 0.55 dex, 
which dominates the $M_\text{BH}$ uncertainty.
We also calculate $M_\text{BH}$ by adopting the latest scaling relation calibrated by \citet{2018ApJ...859..138W}.
Their results are consistent with \citet{2009ApJ...699..800V} but with smaller uncertainties (intrinsic scatter $0.23$ dex).
Table~\ref{tab:mbhledd} shows all $M_\text{BH}$ estimations from two different \ion{Mg}{II} line profile fittings.
For comparison purpose, we adopt $M_\text{BH}$ estimated from the relation provided by \citet{2009ApJ...699..800V} in the following.
Note that we follow the same $M_\text{BH}$ measurement method as used by \citetalias{2017ApJ...849...91M} except for the \ion{Mg}{II} line profile fitting.
Both \ion{Mg}{II} profile fittings give the consistent mass $M_\text{BH}\sim 1.4-1.7 \times 10^8M_\odot$,
making it one of the lightest SMBHs among all $z>6$ quasars ($10^8-10^{10}M_\odot$). 
We use these masses, together with estimates of the bolometric luminosity, $L_\text{bol}$, to calculate the Eddington ratios ($L_\text{bol}/L_\text{Edd}$). 
We first use the relation from \citet{2008ApJ...680..169S}, 
which simply scales the 3000\AA~luminosity by a constant factor of 5.15 to get a zeroth-order estimate of the bolometric luminosity, 
giving $L_\text{bol}/L_\text{Edd}\sim 1$ (Table~\ref{tab:mbhledd}). 
Although this is a rough estimate by assuming all quasars have similar SEDs, 
we still perform the analyses for comparison purposes.
In fact, standard accretion disc models relate the bolometric luminosity to the monochromatic luminosity on the 
Rayleigh-Jeans (low energy) tail of the disc such that $\lambda L_\lambda\propto (M_\text{BH} \dot{M})^{2/3}$ 
\citep{2004A&A...426..797C,2011ApJ...728...98D}, giving a non-linear dependance $L_\text{bol}\propto \lambda L_\lambda^{3/2} / M_\text{BH}$. 
However, these models also predict a continuum slope of $\alpha_\lambda=-7/3$, significantly redder than our data. 
Hence we model the accretion flow and derive a more accurate Eddington ratio for this quasar in detail below (see Sec.~\ref{sec:sed}).

\begin{table*}
	\centering
	\caption{The SMBH mass properties.}
	\label{tab:mbhledd}
	\begin{threeparttable}
	\begin{tabular}{lcccccc} 
		\hline
		Fitted model & $\lambda L_\text{3000}$ & \ion{Mg}{II} FWHM & $M_\text{BH}$ \tnote{a} & $L_\text{bol}$ & $L_\text{bol}/L_\text{Edd}$ & $M_\text{BH}$ \tnote{b} \\
		 & (10$^{45}$ erg s$^{-1}$) & (km s$^{-1}$) & ($\times 10^8 M_\odot$) & (10$^{46}$ erg s$^{-1}$) & & ($\times 10^8 M_\odot$) \\
		\hline
		\ion{Mg}{II} Double Gaussian fit & $3.75 \pm 0.09$ & $1767 \pm 74$ & $1.37^{+3.51}_{-0.99}$ & $1.93^{+1.13}_{-0.71} $ & $1.08^{+3.10}_{-0.80}$ & $1.98^{+1.41}_{-0.82}$ \\
		\ion{Mg}{II} Single Gaussian fit & $3.75 \pm 0.09$ & $1973 \pm 55$ & $1.71^{+4.37}_{-1.23}$ & $1.93^{+1.13}_{-0.71} $ & $0.87^{+2.48}_{-0.64}$ & $2.47^{+1.74}_{-1.02}$ \\
		\hline
	\end{tabular}
	\begin{tablenotes}
		\item[a] estimated by using the scaling relation in \citet{2009ApJ...699..800V}
		\item[b] estimated by using the scaling relation in \citet{2018ApJ...859..138W}
	 \end{tablenotes}
	 \end{threeparttable}	
\end{table*}

\subsection{SED modeling of the continuum slope}
\label{sec:sed}

\begin{figure}
	\includegraphics[width=\columnwidth]{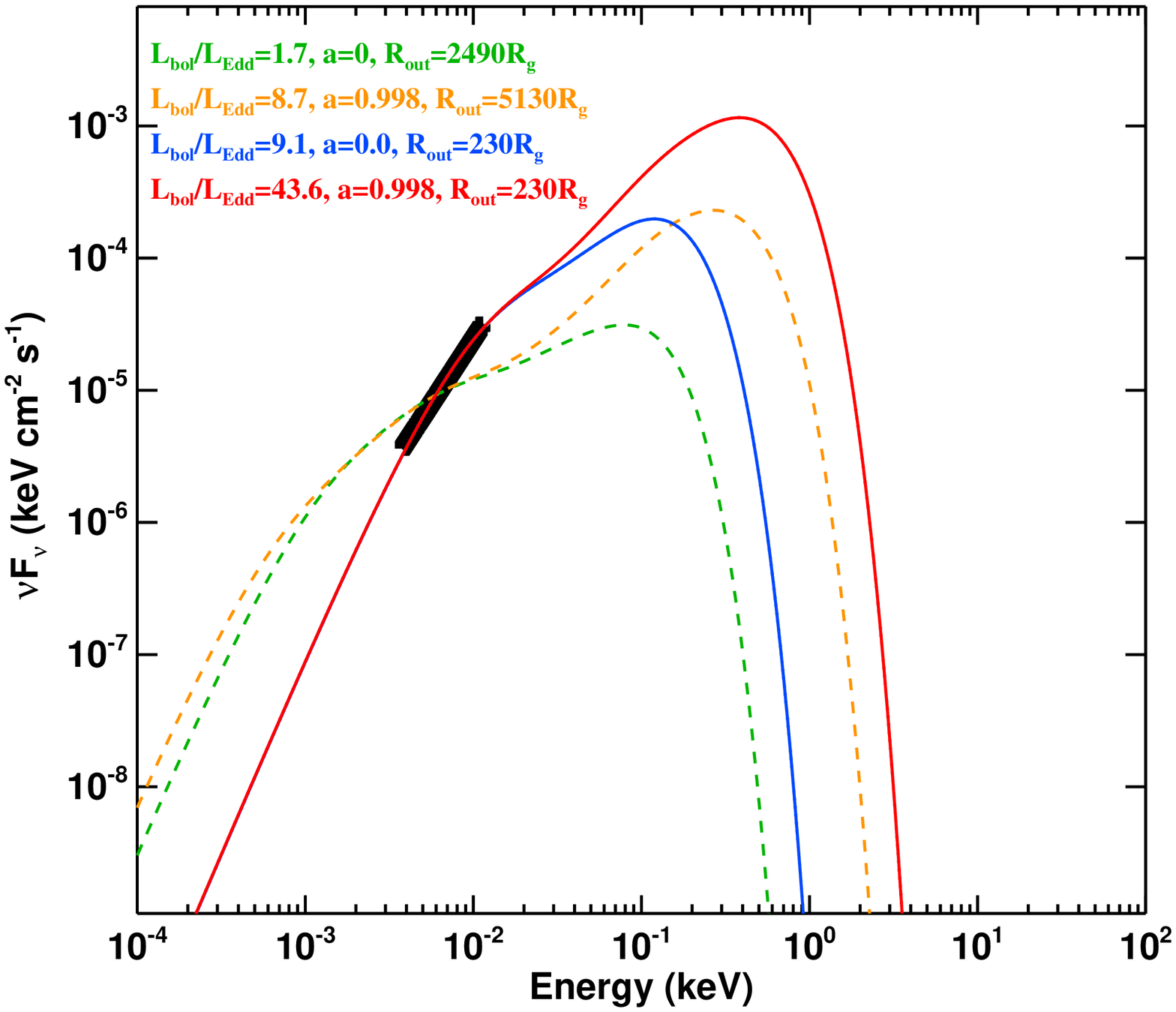}
	\caption{SED fittings using {\sc optxagnf} with parameters in Table~\ref{tab:tab-fit}.
	The heavy black line is the normalisation of the observed UV continuum.
	The orange and green dashed lines use self-gravity radii automatically calculated by {\sc optxagnf} with zero (SED-1) and nearly maximal black hole spins (SED-2), respectively.
	The blue and red solid lines fit outer disc radii as free parameters with zero (SED-3) and nearly maximal black hole spins (SED-4), respectively.\protect\\
	(A colour version of this figure is available in the online journal.)}
	\label{fig:sed_correct}
\end{figure}

We perform SED modeling to investigate the physical mechanisms that produce such a blue continuum, 
and estimate the bolometric luminosity and Eddington ratio.  We use
an energetically self-consistent model based on the standard disc \citep[{\sc optxagnf}:][]{2012MNRAS.420.1848D}, 
where the mass accretion rate is constant with radius, and where the
emissivity is given by the Novikov-Thorne thin disc equation, but where the effect of electron
scattering in the photosphere is taken into account via a colour temperature correction, $f_\text{col}$, such that 
the emission at any radius is modified from a blackbody at the effective temperature ($T_\text{eff}$), 
emitting instead as $B_\nu(f_\text{col}T_\text{eff})/f_\text{col}^4$. 
The colour temperature correction is a function of radius, and
increases markedly with the increase in free electrons at the radius
at which the disc temperature crosses the point at which Hydrogen
changes from being neutral (optical emitting radii) to being ionised
(UV emitting radii). This causes the standard disc continuum in the
rest frame UV as measured here to be redder in the UV than predicted
by a pure blackbody disc model.  Fig.~\ref{fig:sed_correct} shows this for the mean mass
of $1.5 \times 10^8 M_\odot$, for zero and maximal
spin. The normalisation of the observed UV
continuum is determined by the combination of
$(M_\text{BH}\dot{M})^{2/3}$, so gives directly the mass accretion rate
  through the outer disc for our fixed black hole mass, with no spin
  dependence. However, this translates to a 
higher Eddington ratio for high black hole spin due to the increased
  efficiency resulting from the smaller innermost stable circular orbit.
Both high and low spin models have $\alpha_\lambda=-1.7$ in
the rest frame 2000--5000\AA~wavelength range, similar to that
generally observed in quasars \citep{2016A&A...585A..87S,2016ApJ...824...38X},
 but much redder than observed here. 
Without the colour temperature correction, each of these models
would have $\alpha_\lambda=-7/3$ over this wavelength range as it is
dominated by the self similar part of the disc rather than either the
inner (set by black hole spin) or outer (set by self-gravity)
radii. Instead, the data can fit to a much smaller disc, with outer
radius around 230 gravitational radius ($R_\text{g}$) (Fig.~\ref{fig:sed_correct}) 
However, a factor of > 10 reduction in disc size for a
standard disc seems extreme \citep{2010ApJ...724L..59H,2017MNRAS.465..358C},
so we first explore alternative models.

\begin{figure}
	\includegraphics[width=\columnwidth]{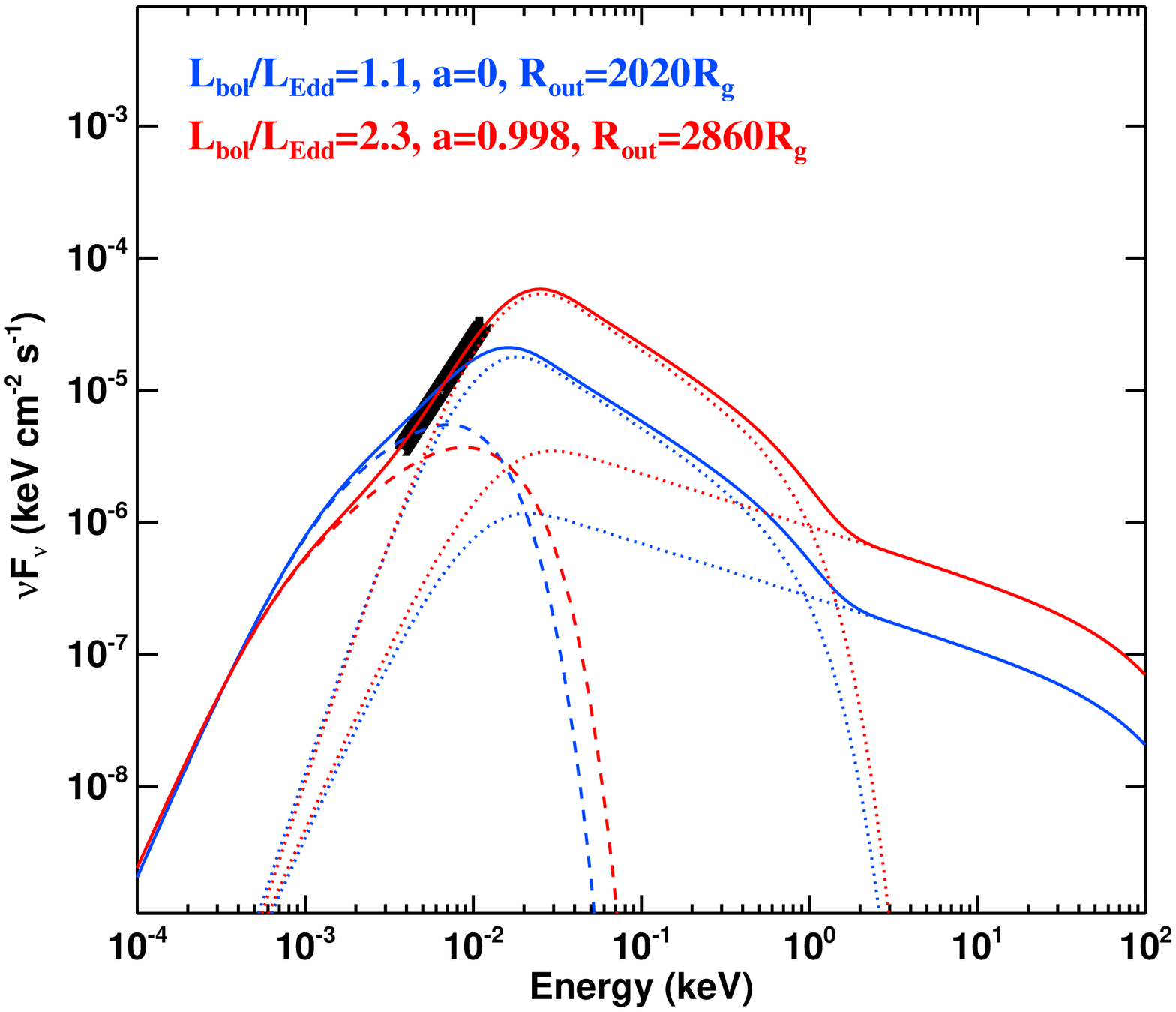}
	\caption{SED fittings with corona X-ray emission with parameters in Table~\ref{tab:tab-fit}.
	The heavy black line is the normalisation of the observed UV continuum.
	The blue and red lines are results with zero (SED-5) and nearly maximal black hole spin (SED-6), respectively.
	The solid, dashed, and dotted lines are total emissions, disc components, and X-ray Comptonisation components, respectively.\protect\\
	(A colour version of this figure is available in the online journal.)}
	\label{fig:sed_wrong}
\end{figure}

The energy conserving code {\sc optxagnf} allows a more complex spectral shape, motivated by the observed SED of
local AGNs. These include additional components as well as the standard disc, with coronal X-ray 
emission from a region which is hot and optically thin forming a high energy power law. 
Below this, AGNs are often observed to have a soft X-ray upturn (the soft X-ray excess) which seems to point back to a UV downturn. 
This can be fit by a second Comptonisation region with very different parameters, where the electrons are warm, and optically thick. 
The energy conserving {\sc optxagnf} model assumes that these components are powered by the accretion energy, so that the disc stops emitting as a (colour temperature corrected) blackbody at some radius $R_\text{cor}$. 
The accretion power emitted within this radius is split between the coronal electrons (fraction $f_\text{pl}$, spectral index $\Gamma$) and the soft X-ray excess warm Comptonisation region characterised by a temperature $kT_\text{e}\sim 0.2$~keV and optical depth $\tau=15$ \citep{2012MNRAS.420.1848D}. 
In Fig.~\ref{fig:sed_wrong},
we fix $f_\text{pl}=0.1$ and $\Gamma=2.4$ as is typical of high $L_\text{bol}/L_\text{Edd}$ AGNs \citep{2017MNRAS.468.3663J} 
and can fit the observed blue slope of the UV spectrum whilst allowing the disc to extend out to its self gravity
radius by making $R_\text{cor}\sim 100R_\text{g}$ so that the transition from the
outer standard blackbody disc to the low temperature, optically thick Comptonisation in this rest frame UV band.
The coronal energy of 10\% is assumed to be contained in the hard X-ray Comptonisation. 
However, as discussed in \citet{2018MNRAS.480.1247K}, the model spectrum is artificially blue at this point as it assumes that the seed photons for the Comptonisation are blackbody
rather than the disc blackbody which is expected for a slab Comptonisation layer above a range of radii in the disc
\citep{2018A&A...611A..59P}. 
Thus these fits are unphysical. 

The only way to fit the very blue observed spectrum with current
models of the accretion flow is with a highly
super-Eddington flow ($L_\text{bol}/L_\text{Edd}\sim 9$ and $44$ for zero and maximal spin, respectively), with small outer extent $R_\text{out}\sim 230R_\text{g}$. 
This small outer radius could indicate that we observe only a limited
section of a larger accretion disc, i.e. that the inner disc is 
puffed up, as expected from a super-Eddington flow \citepalias{1973A&A....24..337S}.
Alternatively, a small outer disc size could instead indicate that the accreting material has very low net angular momentum,
perhaps pointing to the importance of direct infall onto the black hole at these high redshifts. 

\begin{table}
 \centering
   \caption{SED parameters of the {\sc optxagnf} model in Fig.~\ref{fig:sed_correct} and Fig.~\ref{fig:sed_wrong}. 
    $R_\text{cor}$ is the coronal radius. $a$ is the black hole spin. 
   $R_\text{out}$ is the outer radius of the disc. 
   $f$ indicates that the parameter is fixed during the fitting. 
   $s$ indicates the $R_\text{out}$ is equal to the self-gravitational radius. 
   The 90\% confidence range is shown.
   }
    \label{tab:tab-fit}
\begin{tabular}{lccccc}
\hline
Model & $L_\text{bol}/L_\text{Edd}$ & $a$ & $R_\text{cor}$ & $R_\text{out}$ & $\chi^2/dof$ \\
 & & & ($R_\text{g}$) & ($R_\text{g}$) & \\
\hline
SED-1 & 1.7$^{+0.1}_{-0.1}$ & 0$^f$ & -- & 2490$^s$ & 319.0/19 \\
SED-2 & 8.7$^{+0.5}_{-0.5}$ & 0.998$^f$ & -- & 5130$^s$ & 297.4/19 \\
SED-3 & 9.1$^{+2.5}_{-1.7}$ & 0$^f$ & -- & 228$^{+7.0}_{-6.1}$ & 3.7/18 \\
SED-4 & 43.6$^{+11.4}_{-8.3}$ & 0.998$^f$ & -- & 227$^{+7.7}_{-6.3}$ & 3.6/18 \\
SED-5 & 1.1$^{+0.1}_{-0.1}$ & 0$^f$ & 90.1$^{+3.6}_{-3.8}$ & 2020$^s$ & 132.5/18 \\
SED-6 & 2.3$^{+0.9}_{-0.7}$ & 0.998$^f$ & 53.8$^{+1.8}_{-1.7}$ & 2860$^s$ & 1.5/18 \\
\hline
\end{tabular}
\end{table}

\subsection{Chemical abundance in BLR}
\label{sec:metal}

We measure the chemical abundance in the BLR by using the measured emission line flux ratios.
The \ion{Fe}{II}/\ion{Mg}{II} ratio indicator is popular for high redshift quasars since \ion{Mg}{II} and \ion{Fe}{II} can be detected in most of their spectra (Sec.~\ref{sec:metalfe}).
Other metal lines (\ion{N}{V}, \ion{Si}{IV}+\ion{O}{IV]}, and \ion{He}{II}) are fainter and sometimes not detected for high-redshift quasars.
We can measure those faint lines in our spectrum, and we will utilize their ratios to determine the metallicity (Sec.~\ref{sec:metalz}).

\subsubsection{\ion{Fe}{II}/\ion{Mg}{II}}
\label{sec:metalfe}

The \ion{Fe}{II}/\ion{Mg}{II} is a simple chemical abundance tracer as described in Sec.~\ref{sec:intro}.
The method of fitting the \ion{Fe}{II} template and \ion{Mg}{II} line profile is described in Sec.~\ref{sec:fit}.
We integrate the \ion{Fe}{II} template and the double Gaussian function of \ion{Mg}{II} within the rest-frame wavelength range of [2200 -- 3090] \AA\AA~to calculate the \ion{Fe}{II} flux and \ion{Mg}{II} flux, respectively.
The flux error is originated from the uncertainties in the fitting.
We show the \ion{Fe}{II}/\ion{Mg}{II} ratios based on two different \ion{Mg}{II} line profile fittings in Table.~\ref{tab:fe2mg2}.
The comparison between this result and other quasars is discussed in Sec.~\ref{sec:commet}.

\begin{table*}
	\centering
	\caption{\ion{Fe}{II} and \ion{Mg}{II} emission line properties.} 
	\label{tab:fe2mg2}
	\begin{tabular}{lcccc} 
		\hline
		Fitted model & \ion{Mg}{II} Flux & \ion{Fe}{II} Flux & $z_\text{\ion{Mg}{II}}$ & \ion{Fe}{II}/\ion{Mg}{II} \\
		 & (10$^{-16}$ erg s$^{-1}$  cm$^{-2}$) & (10$^{-16}$ erg s$^{-1}$  cm$^{-2}$) & \\
		\hline
		\ion{Mg}{II} Double Gaussian & $1.02 \pm 0.13$ & $1.21 \pm 0.18$ & $6.615 \pm 0.001$ & $1.19 \pm 0.23$ \\
		\ion{Mg}{II} Single Gaussian & $0.73 \pm 0.03$ & $1.84 \pm 0.11$ & $6.615 \pm 0.001$ & $2.52 \pm 0.18$ \\
		\hline
	\end{tabular}
\end{table*}

\subsubsection{Metallicity}
\label{sec:metalz}

\begin{figure}
	\includegraphics[width=\columnwidth]{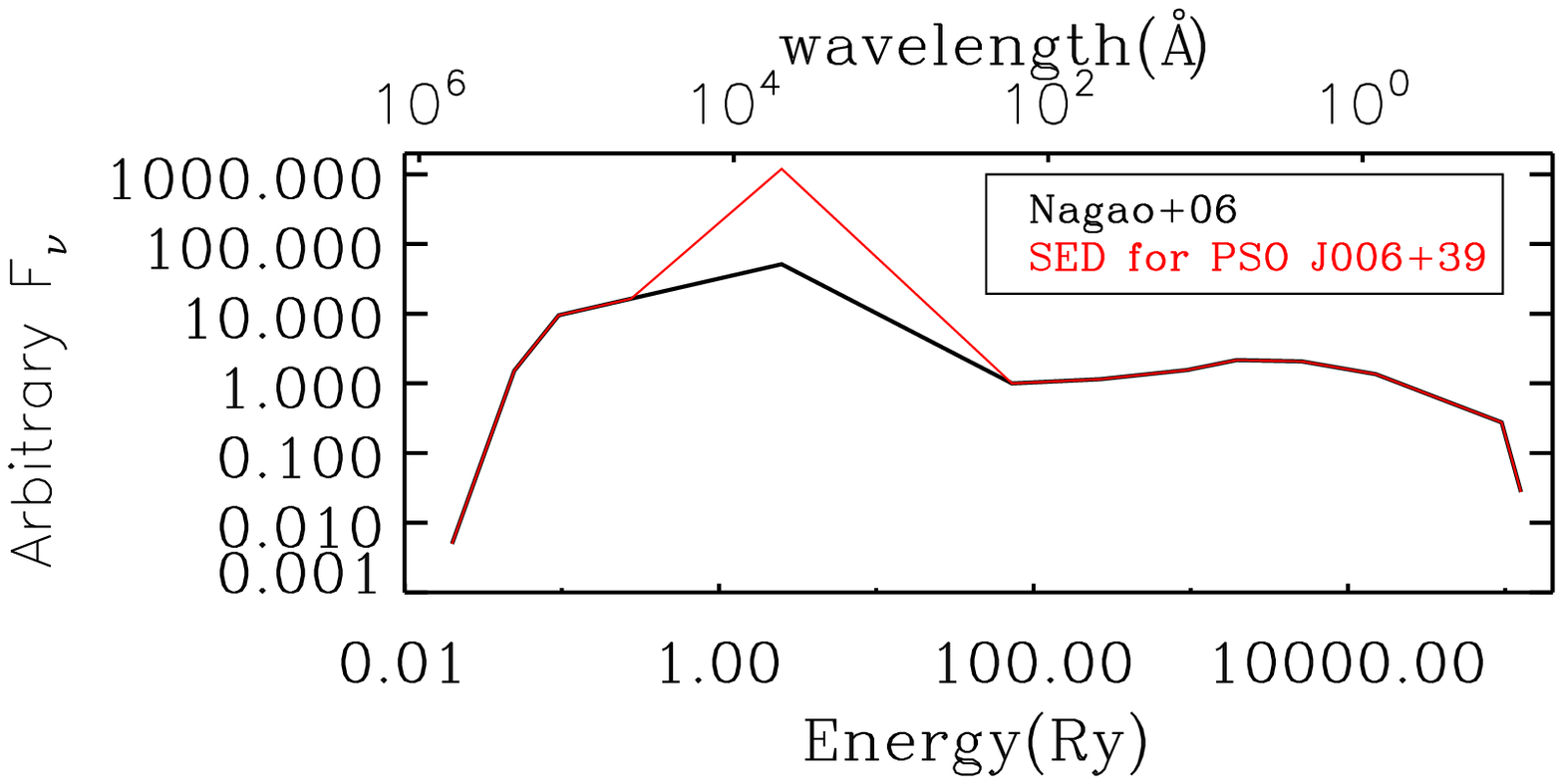}
	\caption{SED of the ionizing source assumed in {\sc Cloudy} model calculation.
	The black line shows the SED with strong UV bump in \citetalias{2006A&A...447..157N}.
	The red line shows the SED corrected for PSO J006+39 in this work.\protect\\
	(A colour version of this figure is available in the online journal.)}
	\label{fig:Cloudysed}
\end{figure}

\begin{figure*}
	\includegraphics[width=\textwidth]{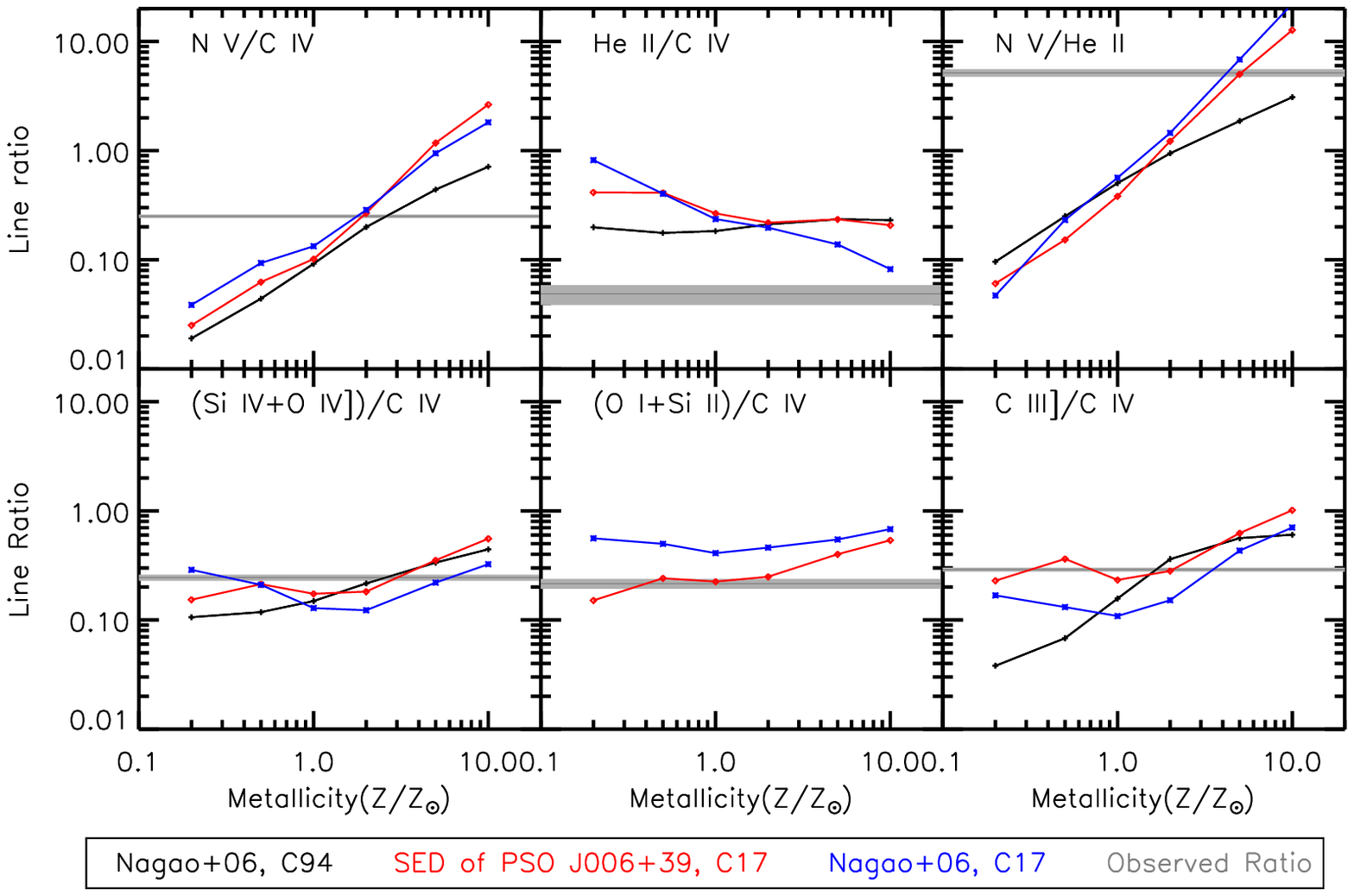}
	\caption{Relations between different line ratios and metallicities.
	The shaded areas show the observed flux ratios of PSO J006+39.
	The black lines show the results from \citetalias{2006A&A...447..157N}.
	The blue lines show the results with the same SED used in \citetalias{2006A&A...447..157N},
	but with the latest {\sc Cloudy} version.\protect\\
	(A colour version of this figure is available in the online journal.)}
	\label{fig:ratio_met}
\end{figure*}

\citet{2006A&A...447..157N} (hereafter \citetalias{2006A&A...447..157N}) investigated the relations between metallicity ($Z$) and several line ratios in the BLR of quasars.
Due to the exceptional blue continuum slope in PSO J006+39,
we reproduce \citetalias{2006A&A...447..157N}'s work with optimization for this quasar.
We adopt the locally optimally emitting cloud (LOC) model \citep{1995ApJ...455L.119B} and apply the latest photoionization code {\sc Cloudy} version 17.01 \citep{2013RMxAA..49..137F}, 
which is newer than {\sc Cloudy} version 94.00 used by \citetalias{2006A&A...447..157N}.
We note that the latest {\sc Cloudy} version 17.01 adopts a newer solar elemental abundance table \citep{1998SSRv...85..161G} than the older version 94.00.
We follow \citetalias{2006A&A...447..157N}'s method to run {\sc Cloudy} calculations by varying gas densities ($n$), ionizing photon fluxes ($\Phi$), and $Z$ in the same ranges.
The metallicities used in the calculations are 0.2, 0.5, 1.0, 2.0, 5.0, and 10.0 $Z/Z_\odot$.
The stopping conditions and the scaling of metal abundances (the secondary Nitrogen abundances) in the calculations are the same as those in \citetalias{2006A&A...447..157N}.
\citetalias{2006A&A...447..157N} assumed two SEDs for AGNs, the typical SED and the typical SED plus a large UV bump, in their {\sc Cloudy} calculations.
We adopt their SED with the large UV bump,
but alter the slope to the same blueness as seen in PSO J006+39 between the lowest rest-frame energy in the observed NIR spectrum and the turning point at 1 Rydberg (Ry) in their SED. 
Fig.~\ref{fig:Cloudysed} shows the difference between the SED with large UV bump in \citetalias{2006A&A...447..157N} and the bluer SED we created.
We integrate the line intensity using equation (2) in \citetalias{2006A&A...447..157N}; i.e.,
\begin{equation}
	L_\text{line}=\iint 4\pi r^2 F_\text{line}(r,n) f(r) g(n) ~\text{dn dr}
	\label{eq:line_int}
\end{equation}
where $f(r)$ and $g(n)$ are the cloud distribution functions for radius ($r$) and $n$.
\citet{1995ApJ...455L.119B} assumed $f(r) \propto r^\Gamma$ and $g(n) \propto n^\beta$.
We follow \citetalias{2006A&A...447..157N} by adopting $\Gamma=-1$ and $\beta=-1$ in the integration of Eq.~\ref{eq:line_int} .
Fig.~\ref{fig:ratio_met} shows the line ratios as a function of metallicities from our {\sc Cloudy} calculations.
We find that the results are different between the same SED but different {\sc Cloudy} versions.
Our results from latest {\sc Cloudy} version show that \ion{N}{V}/\ion{C}{IV} and \ion{N}{V}/\ion{He}{II} line ratios are positively correlated with the metallicity, 
while \ion{He}{II}/\ion{C}{IV} line ratio is negatively correlated with the metallicity.
Besides, (\ion{Si}{IV}+\ion{O}{IV]})/\ion{C}{IV}, (\ion{O}{I}+\ion{Si}{II})/\ion{C}{IV} and \ion{C}{III]}/\ion{C}{IV} line ratios show only weak correlation with metallicity.
By adopting the $\chi^2$ calculations between observed line ratios and predicted ratios from different metallicities, 
our result show $2<Z/Z_\odot<5$ with the smallest $\chi^2$ for $2Z/Z_\odot$.
The interpretation of these results is described in Sec.~\ref{sec:commet}.

\section{Discussion}
\label{sec:dis}

\subsection{Implication of the blue continuum slope}
\label{sec:impslope}

\begin{figure}
	\includegraphics[width=\columnwidth]{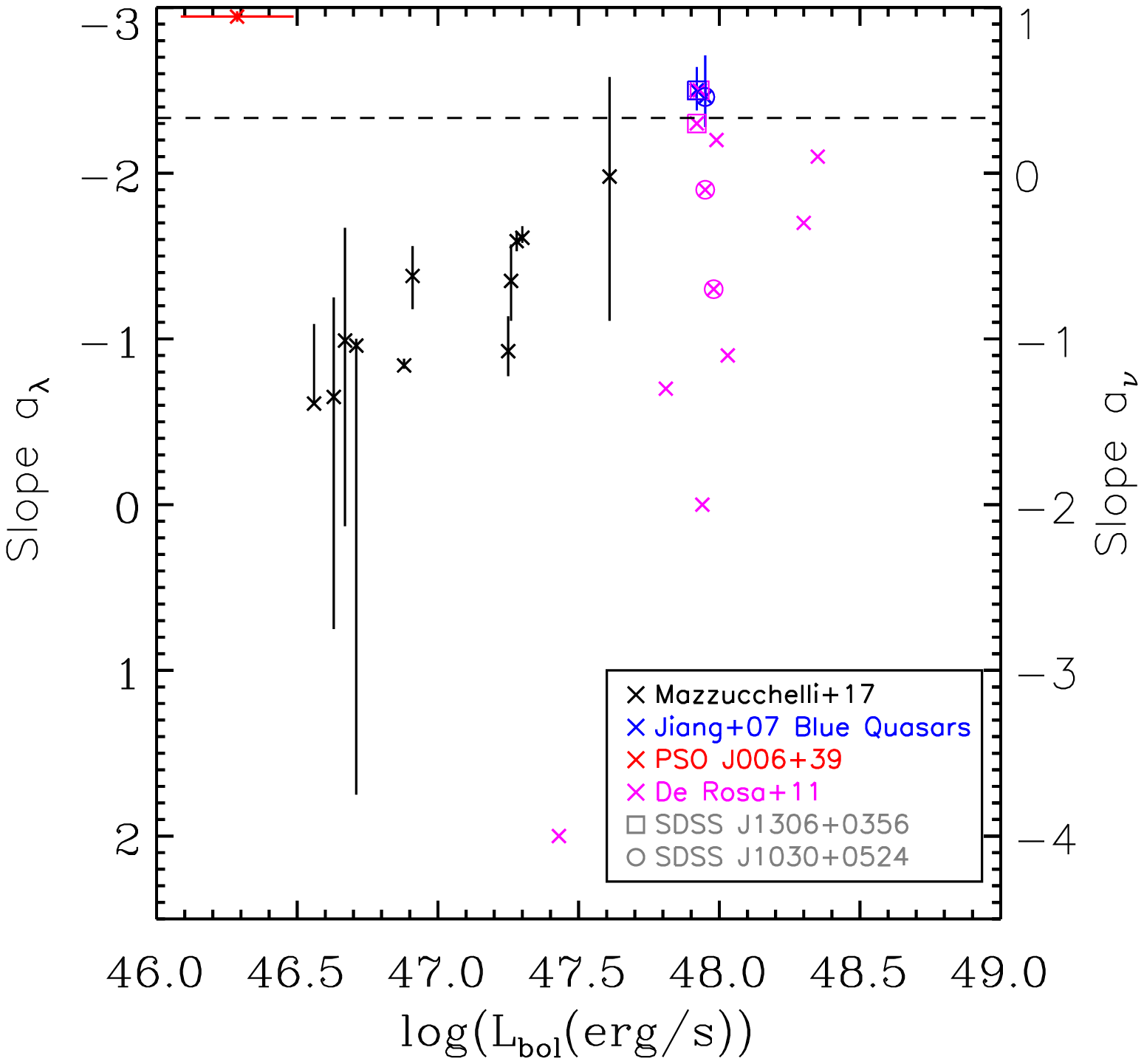}
	\caption{Distribution of the power-law continuum slope versus bolometric luminosity $L_\text{bol}$ for $z>6$ quasars.
	The bolometric luminosity is derived from the scaling relation of \citet{2008ApJ...680..169S} for all the data shown here.
	The horizontal dashed line indicates the prediction from \citetalias{1973A&A....24..337S} model.
	The black, blue and magenta cross-marks show the continuum slope results from \citetalias{2017ApJ...849...91M}, \citet{2007AJ....134.1150J}, and \citet{2011ApJ...739...56D}, respectively.
	The red data with error bar is PSO J006+39 in this work.
	Since the continuum slope of the two blue quasars in \citet{2007AJ....134.1150J} are recalculated in \citet{2011ApJ...739...56D},
	these two quasars in different works are further marked with squares and circles for SDSS J1306+0356 and SDSS J1030+0524, respectively.\protect\\
	(A colour version of this figure is available in the online journal.)}
	\label{fig:qso_conti}
\end{figure}

Fig.~\ref{fig:qso_conti} shows the continuum
slope versus bolometric luminosity ($L_\text{bol}$) estimated from a linear scaling of 3000\AA~luminosity following \citet{2008ApJ...680..169S} for $z > 6$ quasars.
There are only a few objects with continuum slopes bluer than
$\alpha_\lambda=-7/3$ \citep{2007AJ....134.1150J,2011ApJ...739...56D,2014ApJ...790..145D}. 
This is typical of lower redshift quasars also \citep{2016ApJ...824...38X}. 
\citet{2007ApJ...668..682D} show that there are very few quasars around $z=1$ which 
have spectra equivalently as blue as a disc ($\alpha_\nu =+1/3$).
Much of this discrepancy can be due to dust in the host galaxy causing reddening of the quasar 
\citep[e.g.][]{2003AJ....126.1131R,2007ApJ...668..682D}. It may be that dust on large scales in the host galaxy has yet to 
form at the extremely high redshifts probed here, although it is clear that the central regions 
have metallicity close to or beyond solar (Sec.~\ref{sec:metalz}; see also Sec.~\ref{sec:commet}).

Nonetheless, {\it none} of the quasars in the $z\sim 1$ SDSS sample of \citet{2007ApJ...668..682D} have spectra as blue
as PSO J006+39 (equivalent to $\alpha_\nu\sim 1$).
We speculated in Sec.~\ref{sec:sed} that this may be produced if the accretion flow is highly super-Eddington, resulting in the 
inner accretion flow puffing up into a funnel. The apparently small outer radius seen in our spectral fits 
is then set by the outer edge of the funnel rather than by the outer disc itself, and emission from inside the funnel is geometrically beamed by the opening angle of the funnel \citep{2001ApJ...552L.109K}. We note that 
\citet{2014ApJ...797...65W} calculated spectral models of a super-Eddington disc, but ignore the effect of reflection from the funnel, 
so their models remain rather red. 

The uniquely blue spectral index of PSO J006+39 comes then from a combination of
factors, firstly the flow has to be highly super-Eddington in order to
puff up into a funnel, and secondly our line of sight has to intersect
this smaller solid angle. 
The flow is super-Eddington even at low spin ($L_\text{bol}/L_\text{Edd}\sim9$).
However, local, lower mass AGNs which appear similarly super-Eddington do not show such blue spectra
\citep[e.g. RXJ0439 with $\dot{M}/\dot{M}_\text{Edd}=12$ has $\alpha_\lambda=-2.35$,][]{2017MNRAS.471..706J}.
Note that $\dot{M}/\dot{M}_\text{Edd}$ in their work is equivalent to $L_\text{bol}/L_\text{Edd}$ in this work.
This could support an argument for high black hole spin in PSO J006+39 as then the same outer disc mass accretion rate
(which is constrained by the data) leads to a larger bolometric flux so it is more
super-Eddington (e.g. with maximal spin we have $L_\text{bol}/L_\text{Edd}\sim44$). This is more extreme than any local AGN, so would give a 
more extreme accretion geometry.

The very blue slope also has implications for the ionising flux from AGN in the early Universe. 
Simulations predict that the dominant population at $z > 7$ are from low mass black holes accreting at high Eddington fractions \citep[e.g.][]{2011PhDT.......138F}. 
This makes them similar to the local Narrow Line Seyfert 1 (NLS1) class of AGN, and potentially similar to PSO J006+39 but at lower mass. 
The NLS1 in the local Universe have spectra which are much more EUV bright than the standard AGN template spectra such as that shown as the black line in Fig.~\ref{fig:Cloudysed} \citep[see e.g. Fig. 6 in][]{2006ApJ...637..157C}.
Thus their contribution to reionisation of hydrogen/helium in the early universe is severely underestimated by the standard template AGN spectral energy distribution \citep[e.g.][]{2012ApJ...746..125H}. 
However, if the disc puffs up then the escape fraction from the AGN is low, so that their contribution is suppressed. 
More physical models of the AGN spectra and geometry are required in order to properly calculate their contribution to reionisation at high redshift.

Black hole spin plays a large role in the formation and evolution of
SMBHs. Many models assume that the growth rate of the black hole is
limited to $\dot{M}_\text{Edd} = L_\text{Edd}/(\eta(a) c^2) $, 
which is lower for higher black hole spin due to the increased efficiency, $\eta(a)$. 
$c$ is the speed of light.
With this condition it is hard to form massive quasars at
$z=6-7$ from accretion onto low mass black hole seeds in the early
Universe. However, the observation that flows in the local universe
can be super-Eddington shows that it is likely that black holes can
grow at super-Eddington rates with sufficient gas supply \citep{2016MNRAS.455..691J,2017MNRAS.471..706J}.

Recently there have been multiple efforts to understand how black hole
mass accretion rate and spin evolve as the SMBH and their host galaxy
grow from the early Universe to the present day \citep[e.g.][]{2005ApJ...620...69V,2009MNRAS.400..100S,2011MNRAS.410...53F,2018arXiv180608370G}.  
Gas accretion in a fixed plane spins up the
black hole, leading to maximal spin when the black hole mass doubles,
but separate small accretion episodes with random angular momentum can
lead to random walk spin-up/spin down, resulting in low black hole
spin. The supply of gas is linked to the host galaxy and its
evolution, with major mergers leading to widespread star formation,
triggering gas flows into the nucleus to power the accretion fed
growth of the black hole. These also eventually lead to black
hole-black hole mergers, which also affect the spin distribution.
Such models predict a link between black hole spin 
and host galaxy morphology, though they are dependent on multiple
assumptions \citep[e.g.][]{2014ApJ...794..104S,2018arXiv180608370G}.

If the black hole is indeed high spin then we might expect that it can
power a strong jet. 
The Blandford-Znajek (BZ) mechanism \citep{1977MNRAS.179..433B} claims that high black hole spin and the strong magnetic
field are two critical physical properties to produce a powerful jet
in an AGN. Given the high Eddington ratio that we obtain from
the SED fitting in Sec.~\ref{sec:sed}, PSO J006+39 may fit in the scenario
of super-Eddington ratio with slim accretion disk that can produce
jet \citep{2015ASSL..414...45T} if the magnetic field is strong. Currently, we only know a few quasars
that may have jets in the high-redshift universe. In \citet{2015ApJ...804..118B} they reported only nine radio-loud quasars found at $z\sim6$,
corresponding to $6-19\%$ among total high redshift quasars. It is still
unclear that whether PSO J006+39 is radio loud or quiet. There
is no signal from PSO J006+39 in the TIFR GMRT Sky Survey
\citep[TGSS;][]{2017A&A...598A..78I} carried out at the Giant Metrewave
Radio Telescope \citep[GMRT;][]{1991CuSc...60...95S} at 150 MHz. The
detection limit of TGSS is 5 mJy beam$^{-1}$. With an extrapolation
of the slope in the NIR spectrum to 4400\AA, the radio
loudness $R (f_{\nu,\text{1.4GHz}}/f_{\nu,\text{4400\AA}})$ is below $\sim 120$ for PSO J006+39.
Therefore, whether it is a radio-loud quasar still awaits future observations to confirm.

\subsection{BLR's chemical abundance in PSO J006+39}
\label{sec:commet}

\begin{figure}
	\includegraphics[width=\columnwidth]{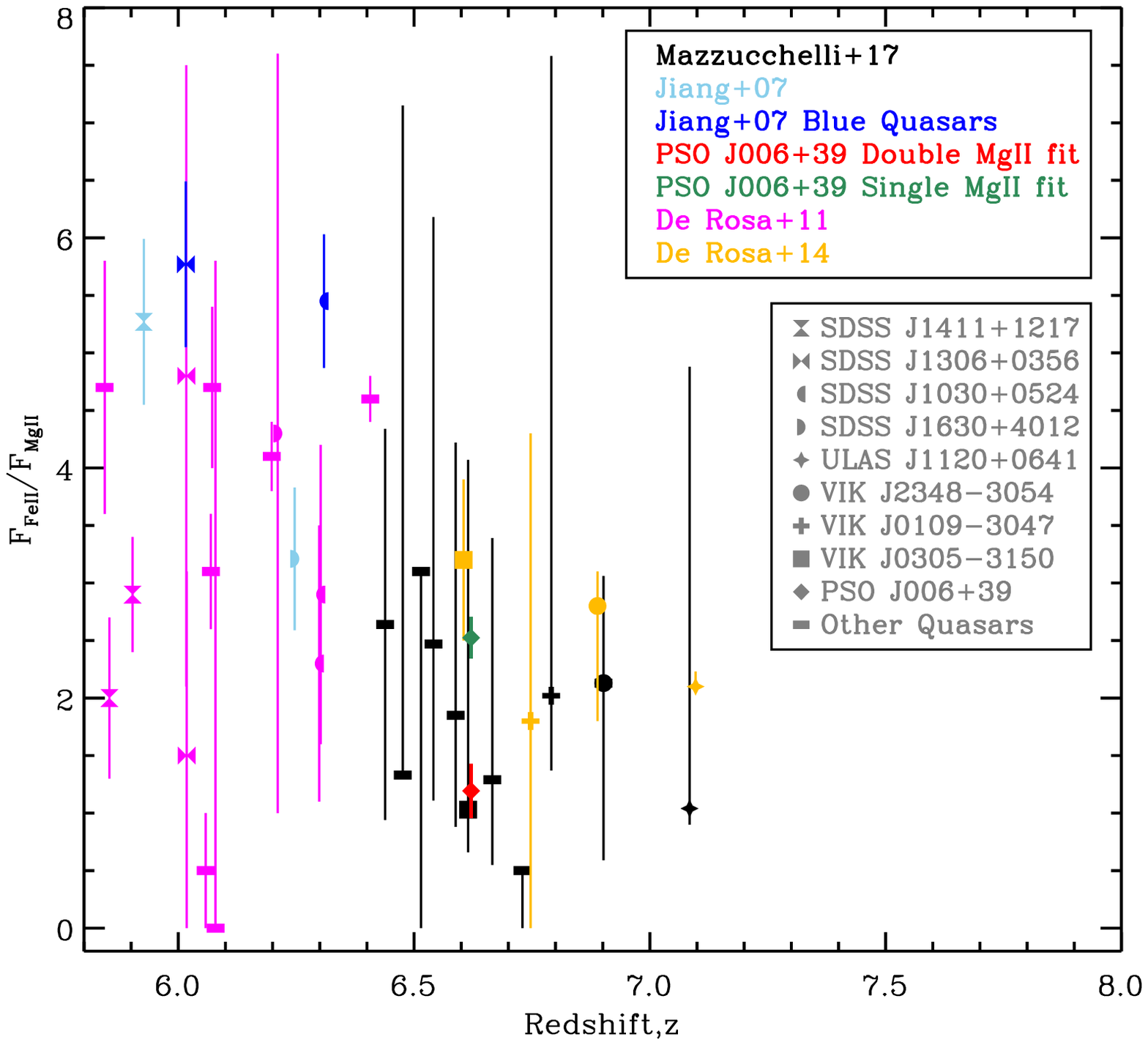}
	\caption{\ion{Fe}{II}/\ion{Mg}{II} line ratio distribution for high redshift quasars.
	The results of high redshift quasars from \citet{2017ApJ...849...91M} and \citet{2011ApJ...739...56D, 2014ApJ...790..145D} are shown in black, magenta, and gold colours, respectively.
	Note that the results done by \citet{2014ApJ...790..145D} are all reprocessed in \citetalias{2017ApJ...849...91M}.
	The sky blue shows the quasars in \citet{2007AJ....134.1150J} with typical continuum slope while the blue shows those with blue continuum slope.
	The circle indicates the line ratio value while the downward triangle indicates the upper limit.
	We show \ion{Fe}{II}/\ion{Mg}{II} line ratio of PSO J006+39 from the double Gaussian fitting for \ion{Mg}{II} in red, 
	but also show the line ratio from the single Gaussian fitting in green for fair comparison with \citetalias{2017ApJ...849...91M}.
	Same quasars with more than one measurement are labeled in the same symbols as indicated in the legend.
	Other quasars with only one measurement are labeled as filled laying bar.\protect\\
	(A colour version of this figure is available in the online journal.)}
	\label{fig:fe2mg2com}
\end{figure}

\begin{figure*}
	\includegraphics[width=\textwidth]{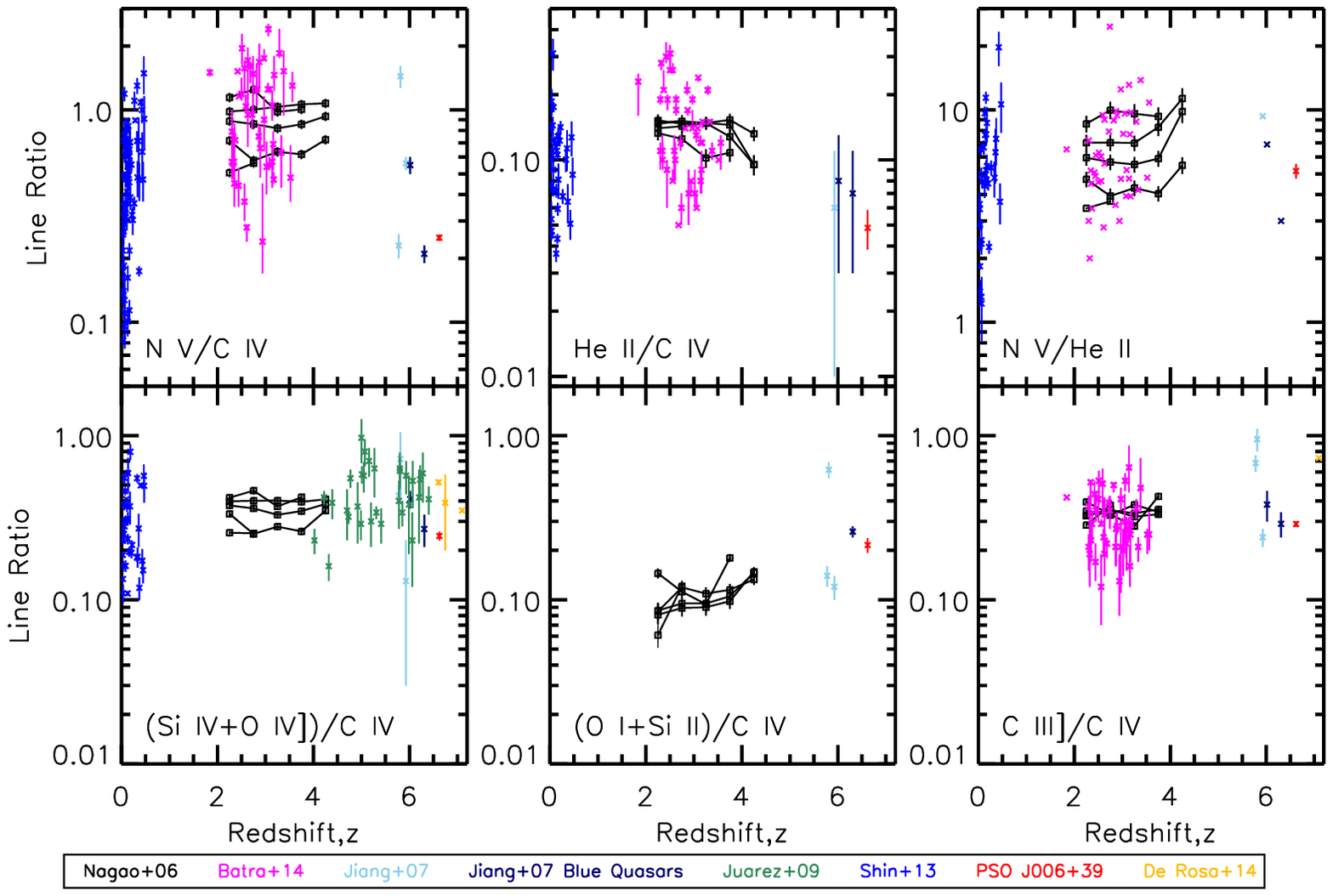}
	\caption{Observed line ratios from low to high redshift quasars.
	The high redshift quasars from \citet{2009A&A...494L..25J}, \citet{2013ApJ...763...58S}, \citet{2014MNRAS.439..771B}, and \citet{2014ApJ...790..145D} are shown in green, blue, magenta, and gold cross-marks with error bars, respectively.
	The sky blue, navy, and red colours are the same as Fig.~\ref{fig:fe2mg2com}.
	Each black line represents one luminosity group from \citetalias{2006A&A...447..157N}, 
	and each luminosity group spans one magnitude.
	The brightest group is $-29.5<M_\text{B}<-28.5$ while the faintest group is $-24.5<M_\text{B}<-25.5$.
	Overall, the brighter group shows higher line ratios except for \ion{He}{II}/\ion{C}{IV}, which has a lower line ratio for brighter groups.\protect\\
	(A colour version of this figure is available in the online journal.)}
	\label{fig:ratio_z}
\end{figure*}

We first examine the chemical abundance of PSO J006+39.
The photoionization calculations show that the actual metallicity is likely $2 < Z/Z_\odot < 5$, while the smallest $\chi^2$ among all metallicities is found at $2 Z_\odot$ (Sec.~\ref{sec:metalz}).
In Fig.~\ref{fig:ratio_met}, most of the observed line ratios fall in this metallicity range except for \ion{He}{II}/\ion{C}{IV} and (\ion{O}{I}+\ion{Si}{II})/\ion{C}{IV} line ratios.
These discrepancies between the line ratios are probably due to the simple assumption of indices ($\Gamma$ and $\beta$) in the line integration (see \citetalias{2006A&A...447..157N} for more discussions).
Although there is ambiguity in the metallicity value,
PSO J006+39 is very likely to have a super solar metallicity in the BLR.
This is consistent with the result from \citet{2007AJ....134.1150J} that $z\gtrsim6$ quasars typically have super solar ($\sim 4Z_\odot$) metallicities.

We now compare PSO J006+39 with other quasars to find out whether it has lower chemical abundance.
Given that there is no strong evidence so far for high redshift quasars showing lower chemical abundance than low redshift quasars (Sec.~\ref{sec:intro}),
finding a low metallicity quasar at high redshift may reveal the reason behind this discrepancy between theoretical prediction and observational evidence.
We show the \ion{Fe}{II}/\ion{Mg}{II} line ratio comparison between PSO J006+39 and quasars at $z\gtrsim5.8$ with available \ion{Fe}{II}/\ion{Mg}{II} line ratio in Fig.~\ref{fig:fe2mg2com}.
We find that PSO J006+39 has relatively low \ion{Fe}{II}/\ion{Mg}{II} line ratio comparing to quasars at $5.8<z<6.4$ if we adopt the result from the double Gaussian fitting for \ion{Mg}{II}.
We caution that this result could be strongly affected by different assumptions in the fittings.
For example,
\citetalias{2017ApJ...849...91M} assumed 10\% of normalization factor for Balmer pseudo-continuum while both \citet{2011ApJ...739...56D} and \citet{2014ApJ...790..145D} assumed 30\%.
We adopt a double Gaussian function for \ion{Mg}{II} line fitting while \citetalias{2017ApJ...849...91M} adopted a single Gaussian function.
\citet{2007AJ....134.1150J} stated that they applied Gaussian convolution on the \ion{Fe}{II} template without showing specific value while \citet{2011ApJ...739...56D} specifically mentioned that they applied FWHM=15\AA.
The small uncertainty in our result may be due to the better SNR in our data comparing to the data used by \citetalias{2017ApJ...849...91M},
but this does not include all these systematic errors.

\begin{figure}
	\includegraphics[width=\columnwidth]{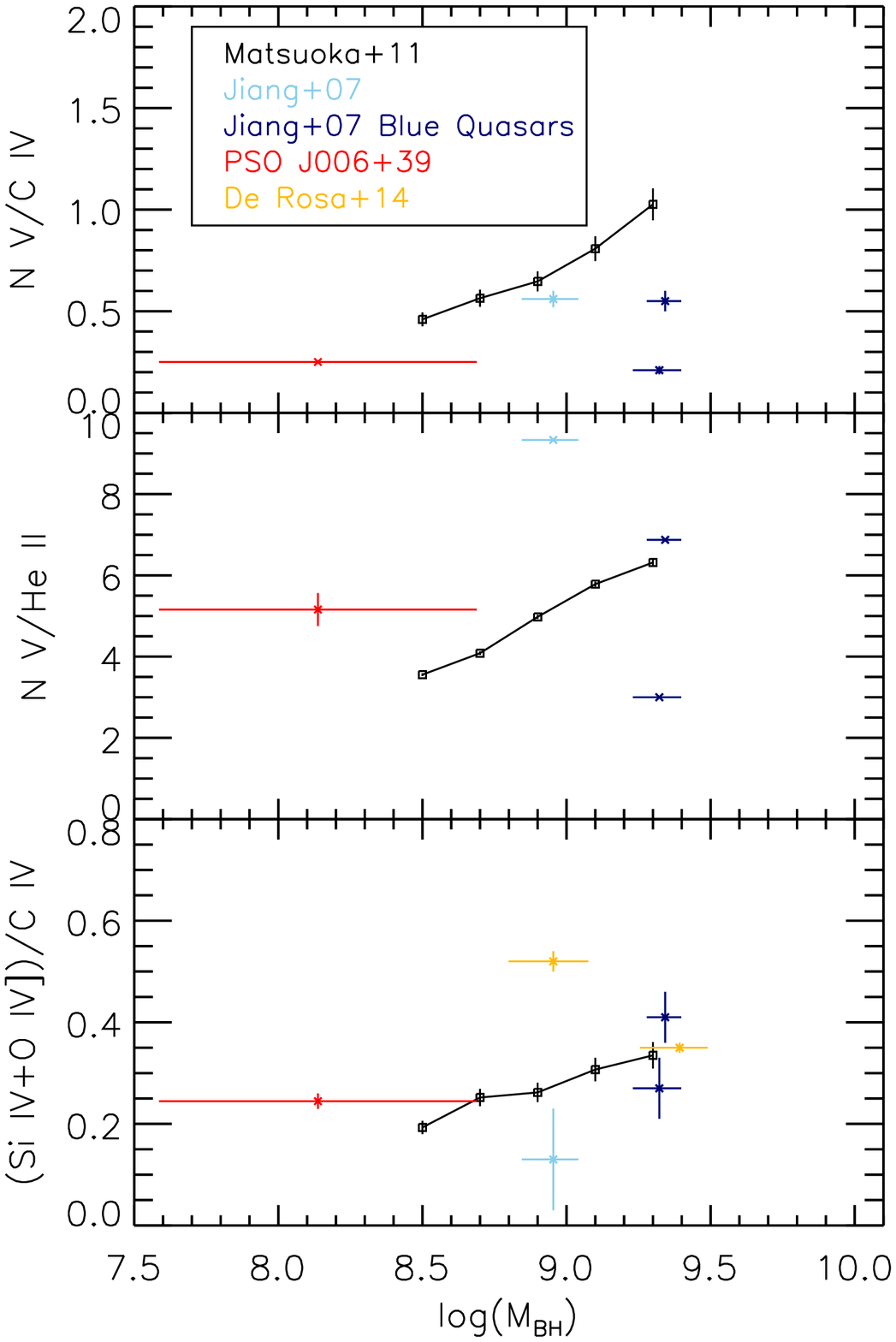}
	\caption{Mass-metallicity relation for three different line ratios.
	Black line shows the group of quasars with $-0.2 \leq$ log($L_\text{bol}/L_\text{Edd}$) $< 0.0$ from \citet{2011A&A...527A.100M}.
	The sky blue, navy, red, and gold cross-marks with error bars are the same in Fig.~\ref{fig:ratio_z}.\protect\\
	(A colour version of this figure is available in the online journal.)}
	\label{fig:ratio_mass}
\end{figure}

We also directly compare individual line ratios between quasars studied so far and PSO J006+39.
This may be useful because \citetalias{2006A&A...447..157N} found that different line ratios have different relations with metallicities.
We show the line ratio as a function of redshift in Fig.~\ref{fig:ratio_z} and as a function of $M_\text{BH}$ in Fig.~\ref{fig:ratio_mass}.
The selected quasar samples include nearby AGNs \citep{2013ApJ...763...58S},
nitrogen-loud SDSS quasars \citep{2014MNRAS.439..771B},
SDSS quasars with NIR spectra \citep{2009A&A...494L..25J},
and $z \sim 6 $ quasars \citep{2007AJ....134.1150J, 2014ApJ...790..145D}.
In addition, the SDSS low redshift ($2.0 < z < 4.5$) quasar sample is classified into different luminosity groups according to \citetalias{2006A&A...447..157N}.
In Fig.~\ref{fig:ratio_z}, we find that both \ion{N}{V}/\ion{C}{IV} and \ion{He}{II}/\ion{C}{IV} line ratios of PSO J006+39 show lower values while other line ratios of this quasar show similar values comparing to quasars from \citetalias{2006A&A...447..157N} and \citet{2014MNRAS.439..771B}.
Other quasar samples are roughly consistent with PSO J006+39.
We also find that the quasars with the UV continuum slopes bluer than the \citetalias{1973A&A....24..337S} model do not show significantly different line ratios comparing to redder quasars in Fig.~\ref{fig:ratio_z}.
This is consistent with the results shown in Fig.~\ref{fig:ratio_met}.
From the comparison in Fig.~\ref{fig:ratio_z}, we find different interpretations for the metallicity of PSO J006+39 using different line ratio indicators.
With the lower value in \ion{N}{V}/\ion{C}{IV} ratio, our {\sc Cloudy} calculation (Fig.~\ref{fig:ratio_met}) suggests that PSO J006+39 indicates lower metallicity comparing to most $z<5$ quasars.
However, Fig.~\ref{fig:ratio_met} suggests that the lower value of PSO J006+39 in \ion{He}{II}/\ion{C}{IV} ratio also indicates slightly higher metallicity comparing to most $z<5$ quasars while the similar value in \ion{N}{V}/\ion{He}{II} has similar metallicity.
It is possible that metal abundance pattern in the BLR of quasars is different from the solar metal abundance pattern. 

We address potential issues in measuring metallicity by using the metal line flux ratios.
First, since luminosity \citepalias{2006A&A...447..157N} and $M_\text{BH}$ \citep[][see also Fig.~\ref{fig:ratio_mass}]{2011A&A...527A.100M} both have relations with the metallicities in the BLR of quasars,
we should select quasars with similar luminosity and $M_\text{BH}$ group for fair metallicity comparison.
However, after extrapolating the power-law continuum of this quasar to optical blue absolute magnitude ($M_\text{B}$), 
we find that the $M_\text{B}$ of PSO J006+39 is fainter ($\sim -24.4$) than any luminosity group presented by \citetalias{2006A&A...447..157N} .
In addition, due to the large uncertainties in the $M_\text{BH}$ measurements, grouping of quasars in terms of similar $M_\text{BH}$ is difficult.
Second, it is doubtful to use the nitrogen lines in metallicity determination for nitrogen-loud quasars \citep[e.g.][]{2002ApJ...564..592H,2008ApJ...679..962J}.
It is possible that strong nitrogen lines are due to the nitrogen overabundance than expected from ordinary scaling relation between N/O and O/H, instead of the overall high metallicity.
Since PSO J006+39 is not a nitrogen-loud quasar (no detection of \ion{N}{IV]}$\lambda1486$ and \ion{N}{III]}$\lambda1750$ and no prominent broad emission line of \ion{N}{V}), 
the lower \ion{N}{V}/\ion{C}{IV} ratio of this quasar comparing to the nitrogen-loud quasars from \citet{2014MNRAS.439..771B} may be caused by their unusual nitrogen abundances.
In this sense, PSO J006+39 can be compared to stacked quasar spectra 
in \citetalias{2006A&A...447..157N} in which the contribution from nitrogen-loud quasars are 
likely small, because nitrogen-loud quasars comprise only 1\% of SDSS 
quasars even though EW of nitrogen emission line in nitrogen-loud 
quasars are one order of magnitude higher than nitrogen quiet \citep{2008ApJ...679..962J}.
In comparison with \citetalias{2006A&A...447..157N}, \ion{N}{V}/\ion{C}{IV} in Fig.~\ref{fig:ratio_z} indicates that the 
metallicity of PSO J006+39 is lower than that of quasars in \citetalias{2006A&A...447..157N} 
in contrast to relatively higher metallicity in \ion{He}{II}/\ion{C}{IV}, when we use the newest 
version of the cloudy. These trends are similar to that we found in comparison with 
nitrogen-loud quasars. This suggests that the potential issue behind the comparison 
in Fig.~\ref{fig:ratio_z} might include not only the unusual nitrogen abundance but also 
the difficulty of accurate modelling of emission lines of BRLs. Actually we found that 
\ion{He}{II}/\ion{C}{IV}-metallicity relation dramatically changed from the correlation to anti-correlation 
depending on the different versions of cloudy as shown in the top-middle panel in Fig.~\ref{fig:ratio_met}.

\section{Conclusions}
\label{sec:con}

We found that PSO J006+39 has a relatively less massive $M_\text{BH}$ ($\sim 10^8 M_\odot$) and an exceptionally blue power-law continuum slope ($\alpha_\lambda=-2.94\pm0.03$) compared to other quasars at $z>6.5$.
We performed SED model fitting and find that this blue continuum slope can be explained by the accretion disc model proposed by \citet{2012MNRAS.420.1848D} with a very small disc ($\sim 230 R_\text{g}$).
This is perhaps due to the flow puffing up into a funnel which is most easily explained if the black hole has high spin and the flow is highly super-Eddington ($L_\text{bol}/L_\text{Edd} \sim 44$) according to \citet{2001ApJ...552L.109K}. 
This indicates the high spin at such high redshift for the first time, 
and appears consistent with the models of black hole spin evolution proposed by \citet{2014ApJ...794..104S}.
Our analysis showed that this quasar exhibits super solar metallicity based on our photoionization calculations, 
but the measured \ion{Fe}{II}/\ion{Mg}{II} line ratio is relatively low compared to known $5.8<z<6.4$ quasars.
However, other different line ratios indicate different metallicities.
Therefore, we could not confirm whether this quasar has lower metallicity compared to other quasars.

\section*{Acknowledgements}

We thank the Gemini staffs in preparing and carrying out the observations.
We thank the anonymous referees for many insightful comments. 
We acknowledge K. Matsuoka, Y. Matsuoka, T. Nagao, M. Nakamura, A. Schulze, X. B. Wu, R. Momose for useful discussions.
TG acknowledges the support by the Ministry of Science and Technology of Taiwan (MoST) through grant 105-2112-M-007-003-MY3, and
YO acknowledges the support by the MoST of Taiwan through grant 106-2112-M-001-008- and 107-2119-M-001-026-.
CJ is supported by the Hundred Talents Program of the Chinese Academy of Sciences.
CJ acknowledges the National Natural Science Foundation of China through grant 11873054.
CD acknowledges the Science and Technology Facilities Council (STFC) 
through grant ST/P000541/1 for support.
EKE acknowledges a post-doctoral fellowship from TUBITAK-BIDEB through 2218 program.

%%%%%%%%%%%%%%%%%%%%%%%%%%%%%%%%%%%%%%%%%%%%%%%%%%



\bibliographystyle{mnras}
\bibliography{example} % if your bibtex file is called example.bib

\begin{thebibliography}{99}
\bibitem[\protect\citeauthoryear{Baldwin et al.}{1995}]{1995ApJ...455L.119B} 
Baldwin J., Ferland G., Korista K., Verner D., 1995, ApJ, 455, L119 
\bibitem[\protect\citeauthoryear{Ba{\~n}ados et al.}{2015}]{2015ApJ...804..118B} 
Ba{\~n}ados E., et al., 2015, ApJ, 804, 118 
\bibitem[\protect\citeauthoryear{Ba{\~n}ados et al.}{2016}]{2016ApJS..227...11B} 
Ba{\~n}ados E., et al., 2016, ApJS, 227, 11 
\bibitem[\protect\citeauthoryear{Ba{\~n}ados et al.}{2018}]{2018Natur.553..473B} 
Ba{\~n}ados E., et al., 2018, Natur, 553, 473
\bibitem[\protect\citeauthoryear{Barth et al.}{2003}]{2003ApJ...594L..95B} 
Barth A.~J., Martini P., Nelson C.~H., Ho L.~C., 2003, ApJ, 594, L95 
\bibitem[\protect\citeauthoryear{Batra \& Baldwin}{2014}]{2014MNRAS.439..771B} 
Batra N.~D., Baldwin J.~A., 2014, MNRAS, 439, 771
\bibitem[\protect\citeauthoryear{Bentz et al.}{2013}]{2013ApJ...767..149B} 
Bentz M.~C., et al., 2013, ApJ, 767, 149 
\bibitem[\protect\citeauthoryear{Blandford \& Znajek}{1977}]{1977MNRAS.179..433B} 
Blandford R.~D., Znajek R.~L., 1977, MNRAS, 179, 433 
\bibitem[\protect\citeauthoryear{Campitiello et al.}{2018}]{2018A&A...612A..59C} 
Campitiello S., Ghisellini G., Sbarrato T., Calderone G., 2018, A\&A, 612, A59 
\bibitem[\protect\citeauthoryear{Casebeer, Leighly, \& Baron}{2006}]{2006ApJ...637..157C} 
Casebeer D.~A., Leighly K.~M., Baron E., 2006, ApJ, 637, 157 
\bibitem[\protect\citeauthoryear{Collin \& Kawaguchi}{2004}]{2004A&A...426..797C} 
Collin S., Kawaguchi T., 2004, A\&A, 426, 797 
\bibitem[\protect\citeauthoryear{Collinson et al.}{2017}]{2017MNRAS.465..358C} 
Collinson J.~S., Ward M.~J., Landt H., Done C., Elvis M., McDowell J.~C., 2017, MNRAS, 465, 358
\bibitem[\protect\citeauthoryear{Davis, Woo, \& Blaes}{2007}]{2007ApJ...668..682D} 
Davis S.~W., Woo J.-H., Blaes O.~M., 2007, ApJ, 668, 682 
\bibitem[\protect\citeauthoryear{Davis \& Laor}{2011}]{2011ApJ...728...98D} 
Davis S.~W., Laor A., 2011, ApJ, 728, 98 
\bibitem[\protect\citeauthoryear{De Rosa et al.}{2011}]{2011ApJ...739...56D} 
De Rosa G., Decarli R., Walter F., Fan X., Jiang L., Kurk J., Pasquali A., Rix H.~W., 2011, ApJ, 739, 56
\bibitem[\protect\citeauthoryear{De Rosa et al.}{2014}]{2014ApJ...790..145D} 
De Rosa G., et al., 2014, ApJ, 790, 145
\bibitem[\protect\citeauthoryear{Diamond-Stanic et al.}{2009}]{2009ApJ...699..782D} 
Diamond-Stanic A.~M., et al., 2009, ApJ, 699, 782 
\bibitem[\protect\citeauthoryear{Dietrich et al.}{2003}]{2003ApJ...596..817D} 
Dietrich M., Hamann F., Appenzeller I., Vestergaard M., 2003, ApJ, 596, 817 
\bibitem[\protect\citeauthoryear{Done et al.}{2012}]{2012MNRAS.420.1848D} 
Done C., Davis S.~W., Jin C., Blaes O., Ward M., 2012, MNRAS, 420, 1848
\bibitem[\protect\citeauthoryear{Gallerani et al.}{2010}]{2010A&A...523A..85G} 
Gallerani S., et al., 2010, A\&A, 523, A85 
\bibitem[\protect\citeauthoryear{Grandi}{1982}]{1982ApJ...255...25G} 
Grandi S.~A., 1982, ApJ, 255, 25 
\bibitem[\protect\citeauthoryear{Grevesse \& Sauval}{1998}]{1998SSRv...85..161G} 
Grevesse N., Sauval A.~J., 1998, SSRv, 85, 161 
\bibitem[\protect\citeauthoryear{Griffin et al.}{2018}]{2018arXiv180608370G} 
Griffin A.~J., Lacey C.~G., Gonzalez-Perez V., Lagos C.~d.~P., Baugh C.~M., Fanidakis N., 2018, arXiv, arXiv:1806.08370 
\bibitem[\protect\citeauthoryear{Fanidakis}{2011}]{2011PhDT.......138F} 
Fanidakis N., 2011, PhDT,
\bibitem[\protect\citeauthoryear{Fanidakis et al.}{2011}]{2011MNRAS.410...53F} 
Fanidakis N., Baugh C.~M., Benson A.~J., Bower R.~G., Cole S., Done C., Frenk C.~S., 2011, MNRAS, 410, 53   
\bibitem[\protect\citeauthoryear{Ferland et al.}{2013}]{2013RMxAA..49..137F} 
Ferland G.~J., et al., 2013, RMxAA, 49, 137 
\bibitem[\protect\citeauthoryear{Freudling, Corbin, \& Korista}{2003}]{2003ApJ...587L..67F} 
Freudling W., Corbin M.~R., Korista K.~T., 2003, ApJ, 587, L67 
\bibitem[\protect\citeauthoryear{Haardt \& Madau}{2012}]{2012ApJ...746..125H} 
Haardt F., Madau P., 2012, ApJ, 746, 125 
\bibitem[\protect\citeauthoryear{Hagino et al.}{2017}]{2017MNRAS.468.1442H} 
Hagino K., Done C., Odaka H., Watanabe S., Takahashi T., 2017, MNRAS, 468, 1442 
\bibitem[\protect\citeauthoryear{Hamann et al.}{2002}]{2002ApJ...564..592H} 
Hamann F., Korista K.~T., Ferland G.~J., Warner C., Baldwin J., 2002, ApJ, 564, 592 
\bibitem[\protect\citeauthoryear{Hao et al.}{2010}]{2010ApJ...724L..59H} 
Hao H., et al., 2010, ApJ, 724, L59 
\bibitem[\protect\citeauthoryear{Leighly}{2004}]{2004ApJ...611..125L} 
Leighly K.~M., 2004, ApJ, 611, 125 
\bibitem[\protect\citeauthoryear{Intema et al.}{2017}]{2017A&A...598A..78I} 
Intema H.~T., Jagannathan P., Mooley K.~P., Frail D.~A., 2017, A\&A, 598, A78 
\bibitem[\protect\citeauthoryear{Iwamuro et al.}{2002}]{2002ApJ...565...63I} 
Iwamuro F., Motohara K., Maihara T., Kimura M., Yoshii Y., Doi M., 2002, ApJ, 565, 63 
\bibitem[\protect\citeauthoryear{Iwamuro et al.}{2004}]{2004ApJ...614...69I} 
Iwamuro F., Kimura M., Eto S., Maihara T., Motohara K., Yoshii Y., Doi M., 2004, ApJ, 614, 69
\bibitem[\protect\citeauthoryear{Jiang et al.}{2007}]{2007AJ....134.1150J} 
Jiang L., Fan X., Vestergaard M., Kurk J.~D., Walter F., Kelly B.~C., Strauss M.~A., 2007, AJ, 134, 1150 
\bibitem[\protect\citeauthoryear{Jiang, Fan, \& Vestergaard}{2008}]{2008ApJ...679..962J} 
Jiang L., Fan X., Vestergaard M., 2008, ApJ, 679, 962 
\bibitem[\protect\citeauthoryear{Jin, Done, \& Ward}{2016}]{2016MNRAS.455..691J} 
Jin C., Done C., Ward M., 2016, MNRAS, 455, 691 
\bibitem[\protect\citeauthoryear{Jin, Done, \& Ward}{2017a}]{2017MNRAS.468.3663J} 
Jin C., Done C., Ward M., 2017a, MNRAS, 468, 3663 
\bibitem[\protect\citeauthoryear{Jin et al.}{2017b}]{2017MNRAS.471..706J} 
Jin C., Done C., Ward M., Gardner E., 2017b, MNRAS, 471, 706 
\bibitem[\protect\citeauthoryear{Juarez et al.}{2009}]{2009A&A...494L..25J} 
Juarez Y., Maiolino R., Mujica R., Pedani M., Marinoni S., Nagao T., Marconi A., Oliva E., 2009, A\&A, 494, L25 
\bibitem[\protect\citeauthoryear{Kaiser et al.}{2002}]{2002SPIE.4836..154K} 
Kaiser N., et al., 2002, SPIE, 4836, 154 
\bibitem[\protect\citeauthoryear{Kaiser et al.}{2010}]{2010SPIE.7733E..0EK} 
Kaiser N., et al., 2010, SPIE, 7733, 77330E
\bibitem[\protect\citeauthoryear{King et al.}{2001}]{2001ApJ...552L.109K} 
King A.~R., Davies M.~B., Ward M.~J., Fabbiano G., Elvis M., 2001, ApJ, 552, L109 
\bibitem[\protect\citeauthoryear{Kubota \& Done}{2018}]{2018MNRAS.480.1247K} 
Kubota A., Done C., 2018, MNRAS, 480, 1247 
\bibitem[\protect\citeauthoryear{Maiolino et al.}{2003}]{2003ApJ...596L.155M} 
Maiolino R., Juarez Y., Mujica R., Nagar N.~M., Oliva E., 2003, ApJ, 596, L155 
\bibitem[\protect\citeauthoryear{Matsuoka et al.}{2011}]{2011A&A...527A.100M} 
Matsuoka K., Nagao T., Marconi A., Maiolino R., Taniguchi Y., 2011, A\&A, 527, A100 
\bibitem[\protect\citeauthoryear{Matteucci \& Greggio}{1986}]{1986A&A...154..279M} 
Matteucci F., Greggio L., 1986, A\&A, 154, 279 
\bibitem[\protect\citeauthoryear{Mazzucchelli et al.}{2017}]{2017ApJ...849...91M} 
Mazzucchelli C., et al., 2017, ApJ, 849, 91
\bibitem[\protect\citeauthoryear{McWilliam}{1997}]{1997ARA&A..35..503M} 
McWilliam A., 1997, ARA\&A, 35, 503 
\bibitem[\protect\citeauthoryear{Nagao, Marconi, \& Maiolino}{2006}]{2006A&A...447..157N} 
Nagao T., Marconi A., Maiolino R., 2006, A\&A, 447, 157 
\bibitem[\protect\citeauthoryear{Petrucci et al.}{2018}]{2018A&A...611A..59P} 
Petrucci P.-O., Ursini F., De Rosa A., Bianchi S., Cappi M., Matt G., Dadina M., Malzac J., 2018, A\&A, 611, A59
\bibitem[\protect\citeauthoryear{Richards et al.}{2003}]{2003AJ....126.1131R} 
Richards G.~T., et al., 2003, AJ, 126, 1131 
\bibitem[\protect\citeauthoryear{Richards et al.}{2011}]{2011AJ....141..167R} 
Richards G.~T., et al., 2011, AJ, 141, 167  
\bibitem[\protect\citeauthoryear{Selsing et al.}{2016}]{2016A&A...585A..87S} 
Selsing J., Fynbo J.~P.~U., Christensen L., Krogager J.-K., 2016, A\&A, 585, A87
\bibitem[\protect\citeauthoryear{Sesana et al.}{2014}]{2014ApJ...794..104S} 
Sesana A., Barausse E., Dotti M., Rossi E.~M., 2014, ApJ, 794, 104 
\bibitem[\protect\citeauthoryear{Sijacki, Springel, \& Haehnelt}{2009}]{2009MNRAS.400..100S} 
Sijacki D., Springel V., Haehnelt M.~G., 2009, MNRAS, 400, 100
\bibitem[\protect\citeauthoryear{Shankar et al.}{2016}]{2016ApJ...818L...1S} 
Shankar F., et al., 2016, ApJ, 818, L1 
\bibitem[\protect\citeauthoryear{Shakura \& Sunyaev}{1973}]{1973A&A....24..337S} 
Shakura N.~I., Sunyaev R.~A., 1973, A\&A, 24, 337 
\bibitem[\protect\citeauthoryear{Shen et al.}{2008}]{2008ApJ...680..169S} 
Shen Y., Greene J.~E., Strauss M.~A., Richards G.~T., Schneider D.~P., 2008, ApJ, 680, 169-190 
\bibitem[\protect\citeauthoryear{Shin et al.}{2013}]{2013ApJ...763...58S} 
Shin J., Woo J.-H., Nagao T., Kim S.~C., 2013, ApJ, 763, 58 
\bibitem[\protect\citeauthoryear{Swarup et al.}{1991}]{1991CuSc...60...95S} 
Swarup G., Ananthakrishnan S., Kapahi V.~K., Rao A.~P., Subrahmanya C.~R., Kulkarni V.~K., 1991, CuSc, 60, 95
\bibitem[\protect\citeauthoryear{Tang et al.}{2017}]{2017MNRAS.466.4568T} 
Tang J.-J., et al., 2017, MNRAS, 466, 4568 
\bibitem[\protect\citeauthoryear{Tchekhovskoy}{2015}]{2015ASSL..414...45T} 
Tchekhovskoy A., 2015, ASSL, 414, 45 
\bibitem[\protect\citeauthoryear{Vanden Berk et al.}{2001}]{2001AJ....122..549V} 
Vanden Berk D.~E., et al., 2001, AJ, 122, 549
\bibitem[\protect\citeauthoryear{Venemans et al.}{2016}]{2016ApJ...816...37V} 
Venemans B.~P., Walter F., Zschaechner L., Decarli R., De Rosa G., Findlay J.~R., McMahon R.~G., Sutherland W.~J., 2016, ApJ, 816, 37  
\bibitem[\protect\citeauthoryear{Vestergaard \& Wilkes}{2001}]{2001ApJS..134....1V} 
Vestergaard M., Wilkes B.~J., 2001, ApJS, 134, 1 
\bibitem[\protect\citeauthoryear{Vestergaard \& Osmer}{2009}]{2009ApJ...699..800V} 
Vestergaard M., Osmer P.~S., 2009, ApJ, 699, 800
\bibitem[\protect\citeauthoryear{Volonteri et al.}{2005}]{2005ApJ...620...69V} 
Volonteri M., Madau P., Quataert E., Rees M.~J., 2005, ApJ, 620, 69
\bibitem[\protect\citeauthoryear{Wang et al.}{2014}]{2014ApJ...797...65W} 
Wang J.-M., Qiu J., Du P., Ho L.~C., 2014, ApJ, 797, 65 
\bibitem[\protect\citeauthoryear{Wu et al.}{2015}]{2015Natur.518..512W} 
Wu X.-B., et al., 2015, Natur, 518, 512 
\bibitem[\protect\citeauthoryear{Woo et al.}{2018}]{2018ApJ...859..138W} 
Woo J.-H., Le H.~A.~N., Karouzos M., Park D., Park D., Malkan M.~A., Treu T., Bennert V.~N., 2018, ApJ, 859, 138 
\bibitem[\protect\citeauthoryear{Xie et al.}{2016}]{2016ApJ...824...38X} 
Xie X., Shao Z., Shen S., Liu H., Li L., 2016, ApJ, 824, 38 
\bibitem[\protect\citeauthoryear{Xu et al.}{2018}]{2018MNRAS.480..345X} 
Xu F., Bian F., Shen Y., Zuo W., Fan X., Zhu Z., 2018, MNRAS, 480, 345 
\end{thebibliography}


%%%%%%%%%%%%%%%%%%%%%%%%%%%%%%%%%%%%%%%%%%%%%%%%%%

%%%%%%%%%%%%%%%%% APPENDICES %%%%%%%%%%%%%%%%%%%%%

%%%%%%%%%%%%%%%%%%%%%%%%%%%%%%%%%%%%%%%%%%%%%%%%%%

% Don't change these lines
\bsp	% typesetting comment
\label{lastpage}
\end{document}